\documentclass{article}%
\usepackage{amsfonts}
\usepackage{aip}
\usepackage{amsmath}
\usepackage{amssymb}
\usepackage{graphicx}%
\providecommand{\U}[1]{\protect\rule{.1in}{.1in}}

\begin{document}

\title{On the Onsager--Wilson Theory of Wien Effect on Strong Binary Electrolytes in
a High External Electric Field}
\author{Byung Chan Eu\\Department of Chemistry, McGill University,\\801 Sherbrooke St. West, Montreal, QC\\H3A 2K6 Canada}
\maketitle

\begin{abstract}
In this review paper, we present and critically re-examine the formal
expressions for the electrophoretic effect and the ionic field appearing in
the unpublished Yale University PhD dissertation of W. S. Wilson which form
the basis of the Onsager--Wilson theory of the Wien effect in the binary
strong electrolyte solutions. It is pointed out that some of the integrals
that make up the flow velocity formula obtained in the thesis and he evaluated
at the position of the center ion in the ionic atmosphere (i.e., the
coordinate origin) diverge. Therefore they cannot be evaluated by means of
contour integrals in the manner performed in his thesis for the reason pointed
out in the text of this paper. In this paper, the results for the integrals in
question are presented, which are alternatively and exactly evaluated. The
details will be described in the follow-up paper presented elsewhere together
with the improved formula for the Wien effect on conductivity.

\end{abstract}

\section{\textbf{Introduction}}%

\setlength{\baselineskip}{20pt}%
Non-Ohmic phenomena in charged media, such as plasmas\cite{plasmas},
semiconductors\cite{semicond}, ionic solutions\cite{ionic}, subjected to high
external electromagnetic fields are much observed in recent years. They not
only have numerous practical applications in science and engineering, but also
pose much challenge theoretically. Especially, in small
systems\cite{micro-nano} confined in space, which consequently are subject to
a very large field gradient, nonlinear phenomena are usually unavoidable and
their theoretical treatment has been of considerable interest as a large body
of literature on the subject matter attests to the interest, relevance, and
activity in the fields. The usual linear theory would not suffice for an
adequate theoretical treatment, and one would require some sort of nonlinear
theory of irreversible processes, whether a kinetic theory (statistical
mechanics approach) or macroscopic irreversible thermodynamics approach is
taken. The present author has developed a nonlinear theory of irreversible
processes both in the kinetic theory (molecular theory)
approach\cite{eubk1,eubk2} or the phenomenological irreversible thermodynamics
approach\cite{eubk3}, when the medium (liquids or gases) is electrically
neutral. Naturally, extension of the theories mentioned of neutral media has
been one of the aims of recent research of the present author.

In electro-physical chemistry, non-Ohmic behavior of conductivity was observed
since 1920s. It is known as the Wien effect\cite{wien} in electrolyte
solutions. The theories of electrical conductivity in electrolyte solutions
have been developed by Lars Onsager and his
collaborators\cite{onsager,onsager1,wilson,kim} over a period stretching over
twenty some years beginning from the late 1920s. Onsager's theories elucidate
the important effects underlying the conduction phenomena in charged media
which we may make use of for many conduction phenomena observed in the fields
and matters mentioned of current interest. Despite their significance and
implications there has been very few follow-up studies of his theories by
other authors in which the theories could have been critically reviewed and
analyzed for further development at the basic level. Although, superficially,
they seem dated from the current standpoint it appears there still are many
features and techniques in them from which we could much benefit for the
current new things we daily encounter in the fields mentioned earlier. On the
strength of this line of reasoning and motivated by the possibility of utility
of his theories for new related problems, the present author has been studying
his theories as closely as possible at the basic level. We would like to
report that even an improvement of one of the results on the Wien effect in
binary electrolytes can be achieved that would affect the experimental
interpretations in the past of the conductivity in electrolyte solutions
carried out on the basis of the Wien effect on electrolytes.

In the path breaking paper of ionic conductivity in strong electrolyte
solutions subject\ to an external electric field, Lars Onsager \cite{onsager1}
in 1927 derived a linear conductance formula for strong electrolyte solutions.
To derive the formula he took into consideration the electrophoretic effect
and the ionic field relaxation effect, which were estimated by means of
physical arguments---on the basis of the Stokes law and a relaxation time
argument. The formula derived established the limiting laws of ionic
conduction, which ever since has appeared in textbooks of physical chemistry
and, particularly, physical chemistry of electrolyte solutions. This early
linear theory was followed up by W. S. Wilson\cite{wilson} in his 1936 Yale
PhD thesis under Onsager's supervision in which he obtained the formula for
ionic conductance in binary strong electrolytes in high external electric
fields---the Wien effect, which was experimentally discovered by M.
Wien\cite{wien} in 1920s. His thesis has not been published in a journal, but
a summary and the tables for the functions representing the electrophoresis
effect and the ionic field relaxation time effect have appeared in the
well-known monumental reference book on physical chemistry of electrolyte
solutions by Harned and Owen\cite{harned}. The reason for not publishing
Wilson's results, according to Harned and Owen, was the partial Fourier
transform method used in Wilson's thesis should be replaced by a full Fourier
transform method to be developed later. However, it took two decades for such
a method to appear in 1957. In the intervening period, the aforementioned
tables prepared by Wilson have been used in connection with the Wien effect in
the literature.

When compared with experiment, the Wien effect formula obtained, however, does
not fully account for the observed conductance data on strong electrolytes.
Therefore, the theory should be improved, but an improvement cannot be
achieved unless we know the theory in its intimate detail. Because of the
aforementioned deficiency, the ionic conductance formula for the Wien effect
has been seminal and played an important role in the development of the idea
of the ionic association theory put forward by A. Patterson and his
collaborators\cite{patterson} who have made use of Bjerrum's ionic association
theory\cite{bjerrum} which, coupled with the weak electrolyte theory of
Onsager\cite{onsagerw}, has enabled them to account for the observed
conductance data\cite{annrevphys}. Thus the idea of degree of ionic
association took a place in physical chemistry lore.

Having formulated a nonlinear transport theory---generalized hydrodynamic
theory---of non-electrolytes (neutral liquids and gases) by means of kinetic
theory\cite{eubk1,eubk2} and its phenomenological version\cite{eubk3}, the
present author in the recent years has been investigating application of the
aforementioned generalized hydrodynamic theory to electrolytic conduction
under high external electric fields often encountered in various small
systems\cite{micro-nano} of recent interests in various fields. The natural
course to take therefore has been to learn the Onsager line of theories before
deciding to apply them to phenomena in mind. Unsatisfied by the brief summary
of the theory of Wien effect given by Wilson in the monograph of Harned and
Owen\cite{harned}, the present author decided to have a close look at Wilson's thesis.

In his thesis, by solving the governing equations for the distribution
functions and Poisson equations which were earlier established by Onsager and
Fuoss\cite{onsager1}, Wilson obtains the solutions for the distribution
functions and ionic potentials. Therewith he then obtains a formal expression
for the ionic field. He also solves the Navier--Stokes equation for velocity
by a (partial) Fourier transform method and obtains a formal expression for
the velocity field. These formal expressions (solutions) are made up of
complicated integrals [in fact, cos$\left(  \alpha x\right)  $ and sin$\left(
\alpha x\right)  $ transforms, where $x$ is the axial coordinate and $\alpha
$\ the transform variable] involving zeroth-order Bessel functions of second
kind $K_{0}(\lambda\rho)$ with non-simple argument where $\rho$ is the radial
distance from the central ion in the cylindrical coordinate system employed
and $\lambda$ is an irrational function of $\alpha$, the field strength, and
the Debye length characterizing the ion atmosphere of the electrolyte solution
of interest. Since the physical quantities of interest are those of the
central ion located at the origin of the coordinates (the center of ionic
atmosphere), he elects to take $x=0$ and $\rho=0$ within the integrals before
evaluating them. Then he resorts to contour integration methods and manages to
evaluate the integrals in closed analytic forms. The results give rise to the
well-known formula for the Wien effect, which was later found to give
qualitatively correct results compared with experiments, but not
quantitatively in the high field regime.

The aforementioned step of taking special values of the variables, $x=0$ and
$\rho=0$, is found to give rise to a difficulty because a couple of the
resulting integrals preclude the contour integral method that involves a
contour at infinite radius in the complex plane. In fact, when closely
examined, they diverge as the argument ($\alpha$) tends to infinity on the
positive real axis. Therefore, it is not permissible to take $\rho=0$, which
allows to take the Bessel function $K_{0}(\lambda\rho)$ at $\rho=0$ before
evaluating the integral. This means, mathematically, that the integrals are
not uniformly convergent with respect to $\rho$ because of the characteristics
of $K_{0}(\lambda\rho)$ at $\rho=0$ in the integrals. In the
follow-up\cite{euwien} of this tutorial review article, the integrals
appearing in the formal expressions for the velocity field and ionic field in
Wilson's thesis are evaluated \textit{alternatively and exactly}. The
integrals can be reduced to definite integrals of another zeroth-order Bessel
function of second kind, $I_{0}\left(  \lambda\rho\right)  $, along the
imaginary axis, which may be evaluated term by term in terms of quadratures of
algebraic functions. They are rather easily amenable to numerical integration.
The main purpose of this tutorial article is to introduce to the reader the
solution procedures leading to Wilson's results from the governing equations
for the distribution functions and the Navier--Stokes equation with the local
force field constructed from the solutions of the governing equations. They
are not simple at all, yet not readily available in the main stream scientific
literature easily accessible at present. (Wilson's thesis is only available on
interlibrary loan from Yale University.) Besides, Wilson's dissertation
contains a number of important typographical errors often requiring checking
the whole solution processes to correct them. In view of the importance and
significance of the Wien effect in nonlinear physico-chemical processes in
electrolytes it is hoped to provide the reader with the details of the
Onsager--Wilson theory of electrolytes. This author believes that their theory
should provide significant insights and methodologies for the modern
investigations in electrical conductivities and transport properties of
electrically charged fluids and condensed matter physics.

This article is organized as follows. In Sec. II, in view of the fact that
Wilson's thesis was not published and its details are not available to the
general audience in the mainstream scientific literature it seems appropriate
to provide the essential materials beyond what is presented in the monograph
of Harned and Owen. As a matter of fact, it contains valuable lessons and
insights for us to theoretically and mathematically treat physical phenomena
in electrolytes and plasmas subjected to external electric fields. Thus, for
notational purposes we will first present the governing equations and the
formal solutions of the governing equations. Then the formal solution is
obtained for the Navier-Stokes equation by using the local force field formula
supplied by the aforementioned solutions for the governing equations. The
velocity field and ionic field obtained are formal, being given in terms of
one-dimensional Fourier transforms of zeroth-order Bessel functions mentioned
earlier. We then will discuss the procedure used by Wilson to evaluate them,
for which he uses the method of contour integration and residues. Instead of
his contour integral methods we will use our own contours different from his.
They are more easily comprehensible in our opinion. We will indicate an
alternative method of evaluation of the integrals, but only the results are
presented, the details of which will be dealt with in the regular article by
the present author elsewhere. This method gives rise to finite integrals of
the zeroth-order Bessel functions of second kind, $I_{0}\left(  z\right)  $,
along the imaginary axis in the $\alpha$ plane, which in fact can be expressed
as quadratures of algebraic functions---polynomials. In Sec. IV, Wilson's
results obtained for the velocity field and the ionic field are then used for
the formula for ionic conductance of binary strong electrolytes. Sec. V is for
the concluding remarks.

\section{\textbf{The Governing Equations and Solutions}}

To treat transport phenomena in electrolyte solutions Onsager and
Fuoss\cite{onsager1} proposed a Fokker--Planck type equation for ion pair
correlation functions---probability distribution functions. The distribution
functions are introduced in the following manner.

Let the two volume elements $dV_{1}$ and $dV_{2}$ be at $\mathbf{r}_{2}$ and
$\mathbf{r}_{1}$, respectively, at time(duration) $t$. Let the fraction of
time $t_{j}$ of $t$ be the time duration in which particle $j$ is found in
$dV_{1}$. The probability (density) $n_{j}$ of finding $j$ in $dV_{1}$ during
that time duration is then, by the ergodic hypothesis\cite{hill},%
\[
\frac{t_{j}}{t}=n_{j}dV_{1},
\]
whereas the probability $n_{i}$ of finding $i$ in $dV_{2}$ during that time
duration is%
\[
\frac{t_{i}}{t}=n_{i}dV_{2}.
\]
Now let $t_{ji}$ be the time duration of $t_{j}$ (i.e., the fraction of
$t_{j}$) in which an ion $i$ is found in $dV_{2}$, given an ion $j$ in
$dV_{1}$. The conditional probability $n_{ji}$ of finding an $i$ ion in
$dV_{2}$ is then%
\[
\frac{t_{ji}}{t_{j}}=n_{ji}dV_{2}.
\]
Similarly, in the reversed case,%
\[
\frac{t_{ij}}{t_{i}}=n_{ij}dV_{1}.
\]
Since
\[
\frac{t_{ji}}{t}=\frac{t_{ij}}{t},
\]
it follows
\begin{equation}
n_{j}n_{ji}dV_{1}dV_{2}=n_{i}n_{ij}dV_{1}dV_{2}. \label{f0}%
\end{equation}
We may write
\begin{equation}
f_{ji}\left(  \mathbf{r}_{1},\mathbf{r}_{21}\right)  \equiv n_{j}n_{ji}%
=n_{i}n_{ij}\equiv f_{ij}\left(  \mathbf{r}_{2},\mathbf{r}_{12}\right)  .
\label{f1}%
\end{equation}
Here $\mathbf{r}_{1}$ and $\mathbf{r}_{2}$ are position vectors of ion $j$ and
ion $i$, respectively, in a suitable coordinate system and
\begin{equation}
\mathbf{r\equiv r}_{21}=\mathbf{r}_{2}-\mathbf{r}_{1}=-\mathbf{r}_{12}.
\label{f2}%
\end{equation}
The $f_{ji}\left(  \mathbf{r}_{1},\mathbf{r}_{21}\right)  $ is the probability
of finding ion $i$ at $\mathbf{r}_{2}$ in $dV_{2}$ at distance $\mathbf{r}%
_{21}$ from ion $j$ at $\mathbf{r}_{1}$ in $dV_{1}$, and inversely for
$f_{ij}\left(  \mathbf{r}_{2},\mathbf{r}_{12}\right)  $. These are pair
distribution functions of the particle pair $\left(  i,j\right)  $.

The charge density $\rho_{j}$ at a position $\mathbf{r}$ from ion $j$ is then
given by%
\begin{equation}
\rho_{j}=\sum_{i}n_{ji}e_{i}=\sum_{i}\frac{f_{ji}}{n_{j}}e_{i}, \label{f3}%
\end{equation}
where $e_{i}$ is the charge on ion $i$. The potential due to the $j$ ion and
its atmosphere is denoted $\psi_{j}\left(  \mathbf{r}_{1},\mathbf{r}%
_{21}\right)  $, which obeys the Poisson equation%
\begin{equation}
\mathbf{\nabla\cdot\nabla}\psi_{j}\left(  \mathbf{r}_{1},\mathbf{r}%
_{21}\right)  =-\frac{4\pi}{D}\sum_{i}\frac{f_{ji}\left(  \mathbf{r}%
_{1},\mathbf{r}_{21}\right)  }{n_{j}}e_{i}, \label{f4}%
\end{equation}
where $D$ is the dielectric constant.

Let the velocity of ion $i$ in the neighborhood of ion $j$ be $\mathbf{v}%
_{ji}=\mathbf{v}_{ji}\left(  \mathbf{r}_{1},\mathbf{r}_{21}\right)  $ and
inversely $\mathbf{v}_{ij}=\mathbf{v}_{ij}\left(  \mathbf{r}_{2}%
,\mathbf{r}_{12}\right)  $. With the relative coordinate $\mathbf{r}$
introduced, the position coordinates $\mathbf{r}_{1}$ and $\mathbf{r}_{2}$ can
be suppressed and we may write $\mathbf{v}_{ji}\left(  \mathbf{r}%
_{1},\mathbf{r}_{21}\right)  =\mathbf{v}_{ji}\left(  \mathbf{r}\right)  $ and
$\mathbf{v}_{ij}\left(  \mathbf{r}_{2},\mathbf{r}_{12}\right)  =\mathbf{v}%
_{ij}\left(  -\mathbf{r}\right)  $.

In the Brownian motion model\cite{smoluchowski} it is assumed that%
\begin{align}
\mathbf{v}_{ji}\left(  \mathbf{r}\right)   &  =\mathbf{v}\left(
\mathbf{r}\right)  +\omega_{i}\left(  \mathbf{K}_{ji}-k_{B}T\ln f_{ji}\right)
,\label{f5a}\\
\mathbf{v}_{ij}\left(  -\mathbf{r}\right)   &  =\mathbf{v}\left(
\mathbf{r}\right)  +\omega_{j}\left(  \mathbf{K}_{ij}-k_{B}T\ln f_{ij}\right)
, \label{f5b}%
\end{align}
where $\mathbf{v}\left(  \mathbf{r}\right)  $ is the barycentric velocity, the
total force $\mathbf{K}_{ji}$ is given in terms of the external force
$\mathbf{F}$, and fluctuating forces of Brownian motion%
\[
\mathbf{K}_{ji}=\mathbf{k}_{i}+\text{ fluctuating forces,}%
\]
and $\omega_{i}$ is the mobility of ion $i$ that produces velocity $\omega
_{i}\mathbf{k}_{i}$ when $\mathbf{k}_{i}$ acts on ion $i$. In the relative
coordinates adopted, the probability $f_{ji}\left(  \mathbf{r}_{1}%
,\mathbf{r}_{21}\right)  $---pair distribution function---is assumed to obey
the equation of continuity%
\begin{equation}
-\frac{\partial f_{ji}\left(  \mathbf{r}_{1},\mathbf{r}_{21}\right)
}{\partial t}=\mathbf{\nabla}_{1}\cdot\left[  f_{ji}\mathbf{v}_{ji}\left(
\mathbf{r}_{1},\mathbf{r}_{21}\right)  +f_{ij}\mathbf{v}_{ij}\left(
\mathbf{r}_{2},\mathbf{r}_{12}\right)  \right]  . \label{f5c}%
\end{equation}
With the Brownian motion model (\ref{f5a}) and (\ref{f5b}) a steady state
Fokker--Planck type equation is obtained from Eq. (\ref{f5c}) for $f_{ji}$:
\begin{align}
-\mathbf{\nabla}\cdot\left[  f_{ji}\mathbf{v}_{ji}\left(  \mathbf{r}\right)
-f_{ij}\mathbf{v}_{ij}\left(  -\mathbf{r}\right)  \right]   &  =\mathbf{\nabla
}_{1}\cdot\omega_{j}\left(  f_{ij}\mathbf{K}_{ij}-k_{B}T\mathbf{\nabla}%
_{1}f_{ij}\right) \label{f6}\\
&  +\mathbf{\nabla}_{2}\cdot\omega_{i}\left(  f_{ji}\mathbf{K}_{ji}%
-k_{B}T\mathbf{\nabla}_{2}f_{ji}\right)  .\nonumber
\end{align}
Equations (\ref{f4}) and (\ref{f6}) are the basic steady-state governing
equations for the distribution functions $f_{ji}$ and potentials $\psi_{j}$ in
the Onsager--Fuoss theory of electrolytic transport processes in external
electric fields. They are coupled as will be clear presently. Further
assumptions/approximations are made to Eq. (\ref{f6}).

Since the fluctuating forces making up the total force $\mathbf{K}_{ji}$
mainly originate from ionic interactions within the ionic atmosphere and
between ionic atmospheres of ions $i$ and $j$, they must be getting
contributions from ionic potentials $\psi_{i}$ and $\psi_{j}$. Neglecting the
ionic forces of $O\left(  e^{2}\right)  $ or higher, the total forces in the
Onsager--Fuoss theory $\mathbf{K}_{ji}$ are looked for in the form%
\begin{equation}
\mathbf{K}_{ji}=\mathbf{k}_{i}-e_{i}\mathbf{\nabla}_{i}\psi_{i}^{\prime
}\left(  0\right)  -e_{i}\mathbf{\nabla}_{i}\psi_{j}\left(  \mathbf{r}%
_{1},\mathbf{r}_{21}\right)  , \label{f7}%
\end{equation}
where the prime denotes the nonequilibrium correction to $\psi_{i}$ beyond the
Debye--H\"{u}ckel equilibrium potential $\psi_{j}^{0}$ obeying the
Poisson--Boltzmann equation%
\begin{equation}
\mathbf{\nabla\cdot\nabla}\psi_{j}^{0}=\frac{4\pi}{k_{B}TD}\sum_{i=1}^{s}%
n_{i}e_{i}^{2}\psi_{j}^{0}=\kappa^{2}\psi_{j}^{0}. \label{f8}%
\end{equation}
Here the Debye parameter (inverse Debye length) $\kappa$ is defined by%
\begin{equation}
\kappa=\sqrt{\frac{4\pi}{k_{B}TD}\sum_{i=1}^{s}n_{i}e_{i}^{2}.} \label{f9}%
\end{equation}
Thus we now see that Eqs. (\ref{f4}) and (\ref{f6}) are coupled. We look for
the solutions in the forms%
\begin{align}
\psi_{j}\left(  \mathbf{r}_{1},\mathbf{r}_{21}\right)   &  =\psi_{j}%
^{0}\left(  r\right)  +\psi_{j}^{\prime}\left(  \mathbf{r}_{1},\mathbf{r}%
_{21}\right)  ,\label{f9a}\\
n_{ji}\left(  \mathbf{r}_{1},\mathbf{r}_{21}\right)   &  =n_{ji}^{0}\left(
r\right)  +n_{ji}^{\prime}\left(  \mathbf{r}_{1},\mathbf{r}_{21}\right)  .
\label{f9b}%
\end{align}
On substitution of Eq. (\ref{f7}), (\ref{f9}), and (\ref{f9b}), the
steady-state governing equations for $f_{ji}$ are now given by the equation%
\begin{align}
\mathbf{\nabla\,}\cdot\,\left[  \,\mathbf{v}_{ji}\left(  \mathbf{r}\right)
-\mathbf{v}_{ij}\mathbf{(-\mathbf{r})\,}\right]  f_{ji}=  &  \omega
_{i}\mathbf{k}_{i}\,\cdot\,\mathbf{\nabla}f_{ji}\left(  \mathbf{r}\right)
-\omega_{j}\mathbf{k}_{j}\,\cdot\,\mathbf{\nabla}f_{ji}\left(  \mathbf{r}%
\right) \nonumber\\
&  -\mathbf{\nabla\,}\cdot\,\omega_{i}n_{j}n_{i}\left[  e_{i}\mathbf{\nabla
}\psi_{j}\left(  \mathbf{r}\right)  +\frac{k_{B}T}{n_{j}n_{i}}\mathbf{\nabla
}f_{ji}\left(  \mathbf{r}\right)  \right] \nonumber\\
&  -\mathbf{\nabla\,}\cdot\,\omega_{j}n_{j}n_{i}\left[  e_{j}\mathbf{\nabla
}\psi_{i}\,\left(  \mathbf{-r}\right)  +\frac{k_{B}T}{n_{j}n_{i}%
}\mathbf{\nabla\,}f_{ji}\left(  \mathbf{r}\right)  \right] \label{f10}\\
&  \qquad\qquad\left(  i,j=1,2,\cdots,s\right)  .\nonumber
\end{align}
Note that $\mathbf{\nabla}_{i}\cdot\mathbf{k}_{i}=0$ is made use of in this
equation since in the experiments of interest the external forces
$\mathbf{k}_{i}$ are maintained constant in space. For this equation we have
also made use of the fact that
\begin{align}
f_{ji}-n_{i}n_{j}  &  =O\left(  e_{i}e_{j}\right)  =O(e^{2}),\label{f10l}\\
f_{ji}\left(  \mathbf{r}\right)   &  =f_{ij}\left(  -\mathbf{r}\right)  ,
\label{f10s}%
\end{align}
and neglected the terms containing $\mathbf{\nabla}\psi_{i}^{\prime}(0)$ and
$\mathbf{\nabla}\psi_{j}^{\prime}(0)$ owing to the fact that they are
$O(e^{2})$. This final form of the governing equation for $f_{ji}$ are coupled
to the Poisson equation
\begin{equation}
\mathbf{\nabla}^{2}\psi_{i}\left(  \mathbf{r}\right)  =-\frac{4\pi}{D}%
\sum_{i=1}^{s}\frac{f_{ji}\left(  \mathbf{r}\right)  e_{i}}{n_{j}}.
\label{f11}%
\end{equation}

Wilson solved this set for binary electrolyte solutions on which the external
electric field $X$ acting in the direction of positive $x$ axis:%
\begin{equation}
k_{1}=e_{1}X,\qquad k_{2}=e_{2}X. \label{f12}%
\end{equation}
Written out for a binary electrolyte---in fact, a uni-uni electrolyte, Eq.
(\ref{f10}) and Poisson equations becomes a set of six coupled equations:%
\begin{align}
n^{2}e\left\{  \nabla^{2}\left[  \psi_{1}\left(  \mathbf{r}\right)  +\psi
_{1}\left(  -\mathbf{r}\right)  \right]  \right\}  +2k_{B}Tf_{11}\left(
\mathbf{r}\right)  =0  &  ,\label{f13a}\\
-n^{2}e\left\{  \nabla^{2}\left[  \psi_{2}\left(  \mathbf{r}\right)  +\psi
_{2}\left(  -\mathbf{r}\right)  \right]  \right\}  +2k_{B}Tf_{22}\left(
\mathbf{r}\right)  =0  &  ,\label{f13b}\\
Xe\left(  \omega_{1}+\omega_{2}\right)  \nabla_{x}f_{12}\left(  \mathbf{r}%
\right)  -n^{2}e\left\{  \omega_{1}\nabla^{2}\psi_{2}\left(  -\mathbf{r}%
\right)  -\omega_{2}\nabla^{2}\psi_{1}\left(  -\mathbf{r}\right)  \right\}   &
\nonumber\\
+k_{B}T\left(  \omega_{1}+\omega_{2}\right)  \nabla^{2}f_{12}\left(
\mathbf{r}\right)  =0  &  ,\label{f13c}\\
-Xe\left(  \omega_{2}+\omega_{1}\right)  \nabla_{x}f_{21}\left(
\mathbf{r}\right)  -n^{2}e\left\{  \omega_{1}\nabla^{2}\psi_{2}\left(
-\mathbf{r}\right)  -\omega_{2}\nabla^{2}\psi_{1}\left(  \mathbf{r}\right)
\right\}   & \nonumber\\
+k_{B}T\left(  \omega_{1}+\omega_{2}\right)  \nabla^{2}f_{21}\left(
\mathbf{r}\right)  =0  &  ,\label{f13d}\\
\mathbf{\nabla}^{2}\psi_{1}\left(  \mathbf{r}\right)  =-\frac{4\pi e}%
{Dn}\left[  f_{11}\left(  \mathbf{r}\right)  -f_{12}\left(  \mathbf{r}\right)
\right]   &  ,\label{f14a}\\
\mathbf{\nabla}^{2}\psi_{2}\left(  \mathbf{r}\right)  =-\frac{4\pi e}%
{Dn}\left[  f_{21}\left(  \mathbf{r}\right)  -f_{22}\left(  \mathbf{r}\right)
\right]   &  . \label{f14b}%
\end{align}
Here $\nabla_{x}=\partial/\partial x$. Note that for a binary electrolyte%
\begin{equation}
n_{1}e_{1}+n_{2}e_{2}=0 \label{f14n}%
\end{equation}
by electroneutrality and $n_{1}=n_{2}$. These coupled equations are solved
subject to the boundary conditions. At this point, it will be useful to remind
ourselves that these governing equations, when solved, would provide us with
the electrical perturbations to the pair distribution functions and the
potentials. In Onsager's theory there is no provision for calculation of the
equilibrium part of the pair distribution functions in the absence of the
external field. The equilibrium distribution functions should be calculated
and supplied separately, if necessary. For them we may make use of the modern
theory of equilibrium charged fluids.\cite{coulombfluid}

\section{\textbf{Boundary Conditions}}

Eqs. (\ref{f13a})--(\ref{f14b}) are subject to the following boundary conditions.

\subsection{\textbf{Flow}}

The number of ions $F_{ij}\left(  \Omega\right)  $ leaving the interior of a
region $S$ is%
\begin{align}
-\frac{\partial F_{ji}}{\partial t}  &  =\int_{S}f_{ji}\left(  \mathbf{r}%
\right)  \left\{  \mathbf{e}_{n}\cdot\left[  \mathbf{v}_{ji}\left(
\mathbf{r}\right)  -\mathbf{v}_{ij}\left(  -\mathbf{r}\right)  \right]
\right\}  dS\nonumber\\
&  =\int_{\Omega}\mathbf{\nabla}\cdot f_{ji}\left(  \mathbf{r}\right)  \left[
\mathbf{v}_{ji}\left(  \mathbf{r}\right)  -\mathbf{v}_{ij}\left(
-\mathbf{r}\right)  \right]  d\Omega=0, \label{f15}%
\end{align}
where $\mathbf{e}_{n}$ is the unit vector normal to the surface $S$. It is
equal to zero by the law of mass conservation. Since volume $\Omega$ can be as
small as possible, the limit of $\Omega\rightarrow0$ may be taken:
\begin{equation}
\lim_{\Omega\rightarrow0}\int_{S}f_{ji}\left(  \mathbf{r}\right)
\mathbf{e}_{n}\cdot\left[  \mathbf{v}_{ji}\left(  \mathbf{r}\right)
-\mathbf{v}_{ij}\left(  -\mathbf{r}\right)  \right]  dS=0. \label{f16}%
\end{equation}
This means that the flow field $f_{ji}\left(  \mathbf{r}\right)  \left[
\mathbf{v}_{ji}\left(  \mathbf{r}\right)  -\mathbf{v}_{ij}\left(
-\mathbf{r}\right)  \right]  $ must be sourceless. That is, $f_{ji}\left(
\mathbf{r}\right)  $ must satisfy this condition. According to Eq. (\ref{f10})
this flow condition implies%
\begin{align}
f_{ji}\left(  \mathbf{r}\right)  \left[  \mathbf{v}_{ji}\left(  \mathbf{r}%
\right)  -\mathbf{v}_{ij}\left(  -\mathbf{r}\right)  \right]   &  =\omega
_{i}\left[  \mathbf{k}_{i}f_{ji}\left(  \mathbf{r}\right)  \right]
-\omega_{j}\left[  \mathbf{k}_{j}f_{ij}\left(  -\mathbf{r}\right)  \right]
\nonumber\\
&  \qquad-\left[  \omega_{i}e_{i}n_{i}n_{j}\mathbf{\nabla}\psi_{j}\left(
\mathbf{r}\right)  +\omega_{j}e_{j}n_{i}n_{j}\mathbf{\nabla}\psi_{i}\left(
-\mathbf{r}\right)  \right] \nonumber\\
&  \qquad-k_{B}T\left(  \omega_{i}+\omega_{j}\right)  \mathbf{\nabla}%
f_{ji}\left(  \mathbf{r}\right)  . \label{f17}%
\end{align}
This condition will be made use of later.

\subsection{\textbf{Ionic Fields}}

The space charge within $\Omega$ is%
\begin{equation}
\int_{\Omega}d\Omega\rho_{j}\left(  \mathbf{r}\right)  =-\frac{D}{4\pi}%
\int_{\Omega}d\Omega\mathbf{\nabla\cdot\nabla}\psi_{j}\left(  \mathbf{r}%
\right)  \label{f18}%
\end{equation}
As $\Omega\rightarrow0$, the volume of the ion $j$ at the origin, the
left-hand side of this equation is equal to $e_{j}$ if the charge is located
at the origin of the coordinates, or equal to zero if the charge is not in
$\Omega$. Thus we may write it as%
\begin{equation}
\rho_{j}\left(  \mathbf{r}\right)  =e_{j}\delta\left(  \mathbf{r}\right)  .
\label{f19}%
\end{equation}
Therefore%
\begin{equation}
\mathbf{\nabla\cdot\nabla}\psi_{j}\left(  \mathbf{r}\right)  =-\frac{4\pi
e_{j}}{D}\delta\left(  \mathbf{r}\right)  \label{f20}%
\end{equation}
or%
\begin{equation}
\lim_{\Omega\rightarrow0}\int_{S}dS\cdot\mathbf{\nabla}\psi_{j}\left(
\mathbf{r}\right)  =-\frac{4\pi e_{j}}{D}\delta\label{f21}%
\end{equation}
where%
\begin{align*}
\delta &  =1\qquad\text{if }e_{j}\text{ is at }\mathbf{r}=0\\
&  =0\qquad\text{otherwise.}%
\end{align*}
Since near $\mathbf{r}=0$, the screening effect is equal to zero and hence the
potential is equal to%
\begin{equation}
\psi_{j}=\frac{e_{j}}{Dr} \label{f22}%
\end{equation}
and in the large distances%
\begin{equation}
\lim_{r\rightarrow\infty}\psi_{j}=0. \label{f23}%
\end{equation}
Hence the boundary conditions on $\psi_{j}$ are deduced to be%
\begin{align}
\psi_{j}-\frac{e_{j}}{Dr}  &  <\infty\text{ as }\left\vert \mathbf{r}%
\right\vert \rightarrow0,\label{f24}\\
\psi_{j}\left(  \infty\right)   &  =0. \label{f25}%
\end{align}
If the charge is of finite size $\sigma_{j}$, then this should be modified to%
\begin{align}
\psi_{j}-\frac{e_{j}}{Dr}  &  <\infty\text{ as }\left\vert \mathbf{r}%
\right\vert >\sigma_{j},\label{f24a}\\
\psi_{j}\left(  \infty\right)   &  =0. \label{f25a}%
\end{align}
The Onsager--Wilson theory does not take the case of finite ion sizes into
consideration as should have been.

\subsection{\textbf{Symmetry Conditions for Potentials and Distribution
Functions}}

\subsubsection{\textbf{Potentials}}

Since the distribution functions are even%
\begin{align}
f_{ii}\left(  \mathbf{r}\right)   &  =f_{ii}\left(  -\mathbf{r}\right)
,\qquad f_{jj}\left(  \mathbf{r}\right)  =f_{jj}\left(  -\mathbf{r}\right)
,\label{f26a}\\
f_{ji}\left(  \mathbf{r}\right)   &  =f_{ij}\left(  -\mathbf{r}\right)  ,
\label{f26b}%
\end{align}
it is possible to deduce the symmetry conditions for the potentials. From the
Poisson equations we find%
\begin{align}
\mathbf{\nabla}^{2}\psi_{j}\left(  \mathbf{r}\right)   &  =-\frac{4\pi e_{j}%
}{Dn}\left[  f_{jj}\left(  \mathbf{r}\right)  -f_{ji}\left(  \mathbf{r}%
\right)  \right]  ,\label{f27a}\\
\mathbf{\nabla}^{2}\psi_{i}\left(  -\mathbf{r}\right)   &  =-\frac{4\pi e_{i}%
}{Dn}\left[  f_{ii}\left(  -\mathbf{r}\right)  -f_{ij}\left(  -\mathbf{r}%
\right)  \right]  . \label{f27b}%
\end{align}
Adding the two equations yields%
\begin{equation}
\mathbf{\nabla}^{2}\left[  \psi_{j}\left(  \mathbf{r}\right)  +\psi_{i}\left(
-\mathbf{r}\right)  \right]  =-\frac{4\pi e}{Dn}\left[  z_{j}f_{jj}\left(
\mathbf{r}\right)  +z_{i}f_{ii}\left(  \mathbf{r}\right)  \right]  .
\label{f27}%
\end{equation}
From this follows%
\begin{equation}
\mathbf{\nabla}^{2}\left[  \psi_{j}\left(  -\mathbf{r}\right)  +\psi
_{i}\left(  \mathbf{r}\right)  \right]  =-\frac{4\pi e}{Dn}\left[  z_{j}%
f_{jj}\left(  \mathbf{r}\right)  +z_{i}f_{ii}\left(  \mathbf{r}\right)
\right]  . \label{f28}%
\end{equation}
Here $e_{i}=ez_{i}$ and $e_{j}=ez_{j}$. Subtracting it from the previous
equation, we obtain%
\begin{equation}
\mathbf{\nabla}^{2}\left[  \psi_{j}\left(  \mathbf{r}\right)  -\psi_{j}\left(
-\mathbf{r}\right)  +\psi_{i}\left(  -\mathbf{r}\right)  -\psi_{i}\left(
\mathbf{r}\right)  \right]  =0. \label{f29}%
\end{equation}
By the boundary conditions on the potentials%
\begin{equation}
\psi_{j}\left(  \mathbf{r}\right)  -\psi_{j}\left(  -\mathbf{r}\right)
<\infty. \label{f30}%
\end{equation}
That is, it is finite. Therefore $\psi_{j}\left(  \mathbf{r}\right)  -\psi
_{j}\left(  -\mathbf{r}\right)  +\psi_{i}\left(  -\mathbf{r}\right)  -\psi
_{i}\left(  \mathbf{r}\right)  $ is also finite. According to the theory of
harmonic analysis, any finite function satisfying the Laplace equation is a
constant. Therefore, we may set%
\[
\psi_{j}\left(  \mathbf{r}\right)  -\psi_{j}\left(  -\mathbf{r}\right)
+\psi_{i}\left(  -\mathbf{r}\right)  -\psi_{i}\left(  \mathbf{r}\right)  =0
\]
or%
\begin{equation}
\psi_{j}\left(  \mathbf{r}\right)  -\psi_{j}\left(  -\mathbf{r}\right)
=\psi_{i}\left(  \mathbf{r}\right)  -\psi_{i}\left(  -\mathbf{r}\right)  .
\label{f31}%
\end{equation}
That is, the odd parts of the potentials are equal \textit{independently of
}$i$\textit{ and }$j$. This means we may set%
\begin{equation}
\psi_{j}\left(  \mathbf{r}\right)  -\psi_{j}\left(  -\mathbf{r}\right)
=\psi_{i}\left(  \mathbf{r}\right)  -\psi_{i}\left(  -\mathbf{r}\right)
=2Y\left(  \mathbf{r}\right)  , \label{f32}%
\end{equation}
where $Y\left(  \mathbf{r}\right)  $ is the odd part of the potential.

The symmetry conditions for the even part of the potential is obtained as
follows. From Eq. (\ref{f10}) taken to $O\left(  e^{2}\right)  $%
\begin{equation}
-\mathbf{\nabla}\cdot\omega_{i}\left[  ez_{i}n^{2}\mathbf{\nabla}\psi
_{i}\left(  \mathbf{r}\right)  +k_{B}T\mathbf{\nabla}f_{ii}\left(
\mathbf{r}\right)  \right]  =0.\nonumber
\end{equation}
On integrating\ it over $\Omega$ and observing the boundary conditions
discussed earlier, we find%
\begin{equation}
0=\mathbf{\nabla}\left[  -\omega_{i}ez_{i}n^{2}\left[  \psi_{i}\left(
\mathbf{r}\right)  +\psi_{i}\left(  -\mathbf{r}\right)  \right]
-2k_{B}Tf_{ii}\left(  \mathbf{r}\right)  \right]  \label{f33}%
\end{equation}
which means%
\[
-\omega_{i}ez_{i}n^{2}\left[  \psi_{i}\left(  \mathbf{r}\right)  +\psi
_{i}\left(  -\mathbf{r}\right)  \right]  -2\omega_{i}k_{B}Tf_{ii}\left(
\mathbf{r}\right)  =c,
\]
but $\psi_{i}\left(  \mathbf{\infty}\right)  =0$ and $f_{ii}\left(
\infty\right)  =n^{2}$. Therefore $c=n^{2}$ and we find%
\begin{equation}
f_{ii}\left(  \mathbf{r}\right)  =f_{ii}\left(  -\mathbf{r}\right)
=n^{2}-\frac{z_{i}en^{2}}{2k_{B}T}\left[  \psi_{i}\left(  \mathbf{r}\right)
+\psi_{i}\left(  -\mathbf{r}\right)  \right]  \label{f34a}%
\end{equation}
and similarly%
\begin{equation}
f_{jj}\left(  \mathbf{r}\right)  =f_{jj}\left(  -\mathbf{r}\right)
=n^{2}-\frac{z_{j}en^{2}}{2k_{B}T}\left[  \psi_{j}\left(  \mathbf{r}\right)
+\psi_{j}\left(  -\mathbf{r}\right)  \right]  . \label{f34b}%
\end{equation}
Upon use of the Poisson equations (\ref{f27}) and (\ref{f28}), it follows%
\begin{align}
\mathbf{\nabla}^{2}  &  \left[  \psi_{j}\left(  \mathbf{r}\right)  +\psi
_{j}\left(  -\mathbf{r}\right)  +\psi_{i}\left(  \mathbf{r}\right)  +\psi
_{i}\left(  -\mathbf{r}\right)  \right] \nonumber\\
&  \qquad=-\frac{8\pi e}{Dn}\left\{  z_{j}n^{2}-\frac{z_{j}^{2}en^{2}}%
{2k_{B}T}\left[  \psi_{j}\left(  \mathbf{r}\right)  +\psi_{j}\left(
-\mathbf{r}\right)  \right]  \right\} \nonumber\\
&  \qquad-\frac{8\pi e}{Dn}\left\{  z_{i}n^{2}-\frac{z_{i}^{2}en^{2}}{2k_{B}%
T}\left[  \psi_{i}\left(  \mathbf{r}\right)  +\psi_{i}\left(  -\mathbf{r}%
\right)  \right]  \right\} \nonumber\\
&  \qquad=\frac{8\pi e^{2}\left\vert z_{i}z_{j}\right\vert n}{Dk_{B}T}\left[
\psi_{j}\left(  \mathbf{r}\right)  +\psi_{j}\left(  -\mathbf{r}\right)
+\psi_{i}\left(  \mathbf{r}\right)  +\psi_{i}\left(  -\mathbf{r}\right)
\right]  , \label{f35}%
\end{align}
which yields
\begin{equation}
\left(  \mathbf{\nabla\cdot\nabla-}\frac{\kappa^{2}}{2}\right)  \left[
\psi_{j}\left(  \mathbf{r}\right)  +\psi_{j}\left(  -\mathbf{r}\right)
+\psi_{i}\left(  \mathbf{r}\right)  +\psi_{i}\left(  -\mathbf{r}\right)
\right]  =0. \label{f36}%
\end{equation}
Note that $z=\left\vert z_{i}\right\vert =\left\vert z_{j}\right\vert $. By
the boundary conditions%
\[
\left[  \psi_{j}\left(  \mathbf{r}\right)  +\psi_{j}\left(  -\mathbf{r}%
\right)  +\psi_{i}\left(  \mathbf{r}\right)  +\psi_{i}\left(  -\mathbf{r}%
\right)  \right]  =\text{finite}%
\]
or%
\begin{equation}
\psi_{j}\left(  \mathbf{r}\right)  +\psi_{j}\left(  -\mathbf{r}\right)
+\psi_{i}\left(  \mathbf{r}\right)  +\psi_{i}\left(  -\mathbf{r}\right)  =0.
\label{f37}%
\end{equation}
Therefore there exists a function $\Gamma\left(  \mathbf{r}\right)  $
independently of $i$ and $j$ such that%
\begin{equation}
\psi_{j}\left(  \mathbf{r}\right)  +\psi_{j}\left(  -\mathbf{r}\right)
=-\left[  \psi_{i}\left(  \mathbf{r}\right)  +\psi_{i}\left(  -\mathbf{r}%
\right)  \right]  =2\Gamma\left(  \mathbf{r}\right)  . \label{f38}%
\end{equation}
Here $\Gamma\left(  \mathbf{r}\right)  $ is the even part of the potential.
Thus we have now identified the even and odd part of the potential. Therefore
we finally obtain%
\begin{align}
\psi_{j}\left(  \mathbf{r}\right)   &  =-\psi_{i}\left(  -\mathbf{r}\right)
=\Gamma\left(  \mathbf{r}\right)  +Y\left(  \mathbf{r}\right)  ,\label{f39a}\\
\psi_{j}\left(  -\mathbf{r}\right)   &  =-\psi_{i}\left(  \mathbf{r}\right)
=\Gamma\left(  \mathbf{r}\right)  -Y\left(  \mathbf{r}\right)  . \label{f39b}%
\end{align}
These are symmetry conditions for ionic potentials.

\subsubsection{\textbf{Diagonal Distribution Functions}}

With conditions (\ref{f39a}) and (\ref{f39b}) we deduce the diagonal parts of
the distribution function matrix
\begin{align}
f_{ii}\left(  \mathbf{r}\right)   &  =f_{ii}\left(  -\mathbf{r}\right)
=n^{2}+\frac{z_{i}en^{2}}{2k_{B}T}\left[  \psi_{i}\left(  \mathbf{r}\right)
+\psi_{i}\left(  -\mathbf{r}\right)  \right]  =n^{2}+\frac{z_{i}en^{2}}%
{k_{B}T}\Gamma\left(  \mathbf{r}\right)  ,\label{f40a}\\
f_{jj}\left(  \mathbf{r}\right)   &  =f_{jj}\left(  -\mathbf{r}\right)
=n^{2}-\frac{z_{j}en^{2}}{2k_{B}T}\left[  \psi_{j}\left(  \mathbf{r}\right)
+\psi_{j}\left(  -\mathbf{r}\right)  \right]  =n^{2}-\frac{z_{j}en^{2}}%
{k_{B}T}\Gamma\left(  \mathbf{r}\right)  . \label{f40b}%
\end{align}

\subsubsection{\textbf{Cross Distribution Functions}}

It is expected from Eq. (\ref{f17}) that cross distribution functions also
consist of even and odd parts. We will write them as%
\begin{equation}
f_{ij}\left(  \mp\mathbf{r}\right)  =f_{ji}\left(  \pm\mathbf{r}\right)
=n^{2}+G(\mathbf{r)\pm}U\left(  \mathbf{r}\right)  . \label{f41}%
\end{equation}
Before proceeding further we would like to note that in Harned and
Owen\cite{harned} $n^{2}$ is inserted in the even part whereas in Wilson's
thesis the factor $n^{2}$ is absent. If $f_{ij}\sim n^{2}$ as $r\rightarrow
\infty$, then the boundary value of $G(\mathbf{r})$ would be $G(\mathbf{r}%
)\rightarrow n^{2}$. If $f_{ij}$ is a perturbation to the equilibrium
distribution function as it is here, there is no need for addition of $n^{2}$.
We have put $n^{2}$ in Eq. (\ref{f41}) assuming it is for the entire
distribution function.

\subsection{\textbf{Differential Equations for Even and Odd Parts}}

Inserting Eq. (\ref{f41}) into Eq. (\ref{f17}), separating the symmetric and
antisymmetric parts, and making use of the Poisson equations, we finally
obtain four coupled second-order differential equations for $G$, $U$, $\Gamma
$, and $Y$:%
\begin{align}
\left(  \mathbf{\nabla}^{2}-\frac{1}{2}\kappa^{2}\right)  G(\mathbf{r}%
)-\frac{1}{2}\eta^{\prime}n^{2}\kappa^{2}\Gamma\left(  \mathbf{r}\right)   &
=-\mu^{\prime}\frac{\partial U(\mathbf{r)}}{\partial x},\label{f42a}\\
\left(  \mathbf{\nabla}^{2}-\frac{1}{2}\kappa^{2}\right)  U(\mathbf{r})  &
=-\mu^{\prime}\frac{\partial G(\mathbf{r})}{\partial x},\label{f42b}\\
\mathbf{\nabla}^{2}\Gamma\left(  \mathbf{r}\right)  -\frac{\kappa^{2}}%
{2}\Gamma\left(  \mathbf{r}\right)   &  =\frac{\kappa^{2}}{2n^{2}\eta^{\prime
}}G\left(  \mathbf{r}\right)  ,\label{f42c}\\
\mathbf{\nabla}^{2}Y\left(  \mathbf{r}\right)   &  =\frac{\kappa^{2}}%
{2n^{2}\eta^{\prime}}U\left(  \mathbf{r}\right)  , \label{f42d}%
\end{align}
where%
\begin{align}
\mu^{\prime}  &  =\frac{zeX}{k_{B}T},\label{f42e}\\
\eta^{\prime}  &  =\frac{ze}{k_{B}T} \label{f42f}%
\end{align}
with
\[
z=\left\vert z_{i}\right\vert =\left\vert z_{j}\right\vert
\]
for binary electrolytes. Solutions of the equations given above would yield
the potentials and distribution functions with which to calculate average
quantities for the binary electrolyte.

\section{\textbf{Formal Solutions}}

Owing to the fact that the external field is applied in the positive $x$
direction the system is axially symmetric. Therefore we will adopt the
cylindrical coordinate system with the $x$ direction as the axis of the
cylinder. The coordinates are denoted $\left(  x,\rho,\theta\right)  $, where
$\theta$ is the azimuthal angle and $\rho$ is the radial coordinate. In this
coordinates the Laplacian is given by%
\begin{equation}
\nabla^{2}=\frac{\partial^{2}}{\partial x^{2}}+\frac{1}{\rho}\frac{\partial
}{\partial\rho}\rho\frac{\partial}{\partial\rho}+\frac{1}{\rho^{2}}%
\frac{\partial^{2}}{\partial\theta^{2}}. \label{f43l}%
\end{equation}
Because of the axial symmetry, the solutions are independent of angle $\theta
$. The angular derivative in $\nabla^{2}$ therefore can be ignored. The
functions $G$, $U$, $\Gamma$, and $Y$ then depend on $x$ and $\rho$ only. We
define one-dimensional Fourier transforms of these functions as follows:%
\begin{align}
\left(
\begin{array}
[c]{c}%
G\left(  x,\rho,0\right) \\
\Gamma\left(  x,\rho,0\right)
\end{array}
\right)   &  =\frac{2}{\pi}\int_{0}^{\infty}d\alpha\cos\left(  \alpha
x\right)  \left(
\begin{array}
[c]{c}%
g\left(  \alpha,\rho\right) \\
\gamma\left(  \alpha,\rho\right)
\end{array}
\right)  ,\label{f43}\\
\left(
\begin{array}
[c]{c}%
U\left(  x,\rho,0\right) \\
Y\left(  x,\rho,0\right)
\end{array}
\right)   &  =\frac{2}{\pi}\int_{0}^{\infty}d\alpha\sin\left(  \alpha
x\right)  \left(
\begin{array}
[c]{c}%
u\left(  \alpha,\rho\right) \\
y\left(  \alpha,\rho\right)
\end{array}
\right)  , \label{f44}%
\end{align}
where $\alpha$ is the Fourier transform variable. It should be noted that odd
functions $\left(  U,Y\right)  $ are sine transforms whereas the even
functions are cosine transforms. Taking Fourier transforms of Eqs.
(\ref{f42a})--(\ref{f42d}) according to Eqs. (\ref{f43}) and (\ref{f44}), we
obtain differential equations for $g\left(  \alpha\rho\right)  $,
$\gamma\left(  \alpha\rho\right)  $, $u\left(  \alpha\rho\right)  $, and
$y\left(  \alpha\rho\right)  $:%
\begin{align}
\left(  \frac{1}{\rho}\frac{d}{d\rho}\rho\frac{d}{d\rho}-\alpha^{2}%
-\frac{\kappa^{2}}{2}\right)  u(\alpha,\rho)  &  =\mu^{\prime}\alpha
g(\alpha,\rho),\label{f46}\\
\left(  \frac{1}{\rho}\frac{d}{d\rho}\rho\frac{d}{d\rho}-\alpha^{2}%
-\frac{\kappa^{2}}{2}\right)  g(\alpha,\rho)-\frac{1}{2}\eta^{\prime}%
n^{2}\kappa^{2}\gamma(\alpha,\rho)  &  =-\mu^{\prime}\alpha u(\alpha
,\rho),\label{f47}\\
\left(  \frac{1}{\rho}\frac{d}{d\rho}\rho\frac{d}{\partial\rho}-\alpha
^{2}-\frac{\kappa^{2}}{2}\right)  \gamma\left(  \alpha,\rho\right)   &
=\frac{\kappa^{2}}{2n^{2}\eta^{\prime}}g\left(  \alpha,\rho\right)
,\label{f48}\\
\left(  \frac{1}{\rho}\frac{d}{d\rho}\rho\frac{d}{d\rho}-\alpha^{2}\right)
y\left(  \alpha,\rho\right)   &  =\frac{\kappa^{2}}{2n^{2}\eta^{\prime}%
}u\left(  \alpha,\rho\right)  . \label{f49}%
\end{align}
The boundary conditions on potentials are deduced from the boundary conditions
already established:%
\begin{align}
\lim_{\rho\rightarrow0}\rho\frac{\partial\gamma(\alpha,\rho)}{\partial\rho}
&  =-\frac{e}{D},\label{f52a}\\
\lim_{\rho\rightarrow0}\rho\frac{\partial y(\alpha,\rho)}{\partial\rho}  &
=0. \label{f52b}%
\end{align}
This set of equations can be solved more concisely if an equivalent eigenvalue
problem is solved for the set. However, to remain as close as possible to the
original approach, we follow his approach.

Eliminating $u(\alpha,\rho)$ and $\gamma\left(  \alpha,\rho\right)  $ between
Eqs. (\ref{f46}), (\ref{f47}), and (\ref{f48}) we obtain a fourth-order
differential equation%
\begin{align}
\left(  \frac{1}{\rho}\frac{d}{d\rho}\rho\frac{d}{d\rho}-c^{2}\right)
^{2}g\left(  \alpha\rho\right)   &  =\left(  \frac{1}{4}\kappa^{4}-\mu
^{\prime2}\alpha^{2}\right)  g(\alpha,\rho),\label{f50}\\
c^{2}  &  =\alpha^{2}+\frac{1}{2}\kappa^{2}. \label{f51}%
\end{align}
The solution of the homogeneous differential equation of Eq. (\ref{f50})
\begin{equation}
\left(  \frac{1}{\rho}\frac{d}{d\rho}\rho\frac{d}{d\rho}-c^{2}\right)  \phi=0,
\label{f53}%
\end{equation}
which is finite at $\rho=\infty$, is found to be the zeroth-order Bessel
function\cite{watson} of second kind $K_{0}\left(  c\rho\right)  $:%
\begin{equation}
\phi\left(  \rho\right)  =\phi_{0}K_{0}\left(  c\rho\right)  . \label{f54}%
\end{equation}
Therefore the solution of Eq. (\ref{f50}), $g(\alpha,\rho)$, must be a linear
combination of zeroth-order Bessel functions of second kind. The coefficients
are determined as follows: Set%
\begin{equation}
g(\alpha,\rho)=\omega\left(  \alpha\right)  K_{0}\left(  \lambda\rho\right)  ,
\label{f55}%
\end{equation}
where $\omega\left(  \alpha\right)  $ and $\lambda$ are determined such that
$\omega\left(  \alpha\right)  K_{0}\left(  \lambda\rho\right)  $ is a solution
of Eq. (\ref{f50}). On substituting Eq. (\ref{f55}) into Eqs. (\ref{f50}) we
find%
\[
\left(  \lambda^{2}-c^{2}\right)  ^{2}=\left(  \frac{1}{4}\kappa^{4}%
-\mu^{\prime2}\alpha^{2}\right)
\]
and the solution for $\lambda^{2}$ is%
\begin{align}
\lambda_{1}^{2}  &  =c^{2}+\frac{1}{2}\kappa^{2}R=\alpha^{2}+\frac{1}{2}%
\kappa^{2}+\frac{1}{2}\kappa^{2}R,\label{f56a}\\
\lambda_{2}^{2}  &  =c^{2}-\frac{1}{2}\kappa^{2}R=\alpha^{2}+\frac{1}{2}%
\kappa^{2}-\frac{1}{2}\kappa^{2}R, \label{f56b}%
\end{align}
where%
\begin{equation}
R=\sqrt{1-\frac{4\mu^{\prime2}\alpha^{2}}{\kappa^{4}}}. \label{f57}%
\end{equation}
The existence of two values of $\lambda$ implies that the solution for
$g(\alpha,\rho)$ is the linear combination%
\begin{equation}
g(\alpha,\rho)=\omega_{1}\left(  \alpha\right)  K_{0}\left(  \lambda_{1}%
\rho\right)  +\omega_{2}\left(  \alpha\right)  K_{0}\left(  \lambda_{2}%
\rho\right)  . \label{f58}%
\end{equation}
Similarly, the functions $u\left(  \alpha,\rho\right)  $, $\gamma\left(
\alpha,\rho\right)  $, and $y\left(  \alpha,\rho\right)  $ can be determined
as the linear combinations%
\begin{align}
\gamma\left(  \alpha,\rho\right)   &  =\chi_{1}\left(  \alpha\right)
K_{0}\left(  \lambda_{1}\rho\right)  +\chi_{2}\left(  \alpha\right)
K_{0}\left(  \lambda_{2}\rho\right)  +\chi_{3}\left(  \alpha\right)
K_{0}\left(  \lambda_{3}\rho\right)  ,\label{f59}\\
u\left(  \alpha,\rho\right)   &  =\psi_{1}\left(  \alpha\right)  K_{0}\left(
\lambda_{1}\rho\right)  +\psi_{2}\left(  \alpha\right)  K_{0}\left(
\lambda_{2}\rho\right)  +\psi_{3}\left(  \alpha\right)  K_{0}\left(
\lambda_{3}\rho\right)  ,\label{f60}\\
y\left(  \alpha,\rho\right)   &  =\xi_{1}\left(  \alpha\right)  K_{0}\left(
\lambda_{1}\rho\right)  +\xi_{2}\left(  \alpha\right)  K_{0}\left(
\lambda_{2}\rho\right)  +\xi_{3}\left(  \alpha\right)  K_{0}\left(
\lambda_{3}\rho\right) \nonumber\\
&  \qquad+\xi_{4}\left(  \alpha\right)  K_{0}\left(  \lambda_{4}\rho\right)  ,
\label{f61}%
\end{align}
where%
\begin{align}
\lambda_{3}^{2}  &  =c^{2}=\alpha^{2}+\frac{1}{2}\kappa^{2},\label{f62}\\
\lambda_{4}  &  =\alpha. \label{f63}%
\end{align}
The coefficients $\omega_{1}\left(  \alpha\right)  $, $\chi_{1}\left(
\alpha\right)  $, $\psi_{1}\left(  \alpha\right)  $, $\xi_{1}\left(
\alpha\right)  $, etc. are determined on substitution of Eqs. (\ref{f58}%
)--(\ref{f61}) into Eqs. (\ref{f46})--(\ref{f49}). Since $K_{0}\left(
\lambda_{1}\rho\right)  $ and $K_{0}\left(  \lambda_{2}\rho\right)  $,
$K_{0}\left(  \lambda_{3}\rho\right)  $, and $K_{0}\left(  \lambda_{4}%
\rho\right)  $ are independent, the aforementioned substitution yields the set
of equations for the coefficients:%
\begin{align}
\frac{1}{2}\kappa^{2}R\psi_{1}\left(  \alpha\right)  -\mu^{\prime}\alpha
\omega_{1}\left(  \alpha\right)   &  =0,\label{f64a}\\
\frac{1}{2}\kappa^{2}R\psi_{2}\left(  \alpha\right)  +\mu^{\prime}\alpha
\omega_{2}\left(  \alpha\right)   &  =0,\label{f64b}\\
\left(  \lambda_{1}^{2}-\alpha^{2}-\frac{\kappa^{2}}{2}\right)  \omega
_{1}\left(  \alpha\right)  +\frac{1}{2}\eta^{\prime}n^{2}\kappa^{2}\chi
_{1}\left(  \alpha\right)  +\mu^{\prime}\alpha\psi_{1}\left(  \alpha\right)
&  =0,\label{f65a}\\
\left(  \lambda_{2}^{2}-\alpha^{2}-\frac{\kappa^{2}}{2}\right)  \omega
_{2}\left(  \alpha\right)  +\frac{1}{2}\eta^{\prime}n^{2}\kappa^{2}\chi
_{2}\left(  \alpha\right)  +\mu^{\prime}\alpha\psi_{2}\left(  \alpha\right)
&  =0,\label{f65b}\\
\frac{1}{2}\eta^{\prime}n^{2}\kappa^{2}\chi_{3}\left(  \alpha\right)
-\mu^{\prime}\alpha\psi_{3}\left(  \alpha\right)   &  =0,\label{f65c}\\
\left[  \left(  \lambda_{1}^{2}-\alpha^{2}-\frac{\kappa^{2}}{2}\right)
\chi_{1}\left(  \alpha\right)  -\frac{\kappa^{2}}{2n^{2}\eta^{\prime}}%
\omega_{1}\left(  \alpha\right)  \right]   &  =0,\label{f66a}\\
\left[  \left(  \lambda_{2}^{2}-\alpha^{2}-\frac{\kappa^{2}}{2}\right)
\chi_{2}\left(  \alpha\right)  -\frac{\kappa^{2}}{2n^{2}\eta^{\prime}}%
\omega_{2}\left(  \alpha\right)  \right]   &  =0,\label{f66b}\\
\left(  \lambda_{1}^{2}-\alpha^{2}\right)  \xi_{1}\left(  \alpha\right)
-\frac{\kappa^{2}}{2n^{2}\eta^{\prime}}\psi_{1}\left(  \alpha\right)   &
=0,\label{f67b}\\
\left(  \lambda_{2}^{2}-\alpha^{2}\right)  \xi_{2}\left(  \alpha\right)
-\frac{\kappa^{2}}{2n^{2}\eta^{\prime}}\psi_{2}\left(  \alpha\right)   &
=0,\label{f67c}\\
\left(  \lambda_{3}^{2}-\alpha^{2}\right)  \xi_{3}\left(  \alpha\right)
-\frac{\kappa^{2}}{2n^{2}\eta^{\prime}}\psi_{3}\left(  \alpha\right)   &  =0.
\label{f67a}%
\end{align}

From the boundary conditions follow the equations%
\begin{align}
\sum_{l=1}^{3}\chi_{l}\left(  \alpha\right)   &  =\frac{e}{D},\label{f68a}\\
\sum_{l=1}^{3}\xi_{l}\left(  \alpha\right)   &  =0, \label{f68b}%
\end{align}
while since the flow is sourceless and the field is divergenceless it follows
\begin{align*}
\lim_{\rho\rightarrow0}\rho\left[  n^{2}\eta^{\prime}\frac{\partial
\gamma(\alpha,\rho)}{\partial\rho}-\frac{\partial g(\alpha,\rho)}{\partial
\rho}\right]   &  =0,\\
\lim_{\rho\rightarrow0}\rho\frac{\partial u(\alpha,\rho)}{\partial\rho}  &
=0,
\end{align*}
which provide additional equations%
\begin{align}
-n^{2}\eta^{\prime}\left(  \chi_{1}+\chi_{2}+\chi_{3}\right)  +\omega
_{1}+\omega_{2}  &  =0,\label{f69a}\\
\psi_{1}+\psi_{2}+\psi_{3}  &  =0. \label{f69b}%
\end{align}
Therefore on combining Eqs. (\ref{f68a})--(\ref{f69b}) we obtain%
\begin{align}
\omega_{1}+\omega_{2}  &  =\frac{en^{2}\eta^{\prime}}{D},\label{f70}\\
\chi_{1}+\chi_{2}+\chi_{3}  &  =\frac{e}{D}. \label{f71}%
\end{align}
There are 13 equations for 12 variables. These equations are first reduced to
equations for $\chi_{i}$ only. Then all other variables are determined from
$\chi_{i}$. We obtain%
\begin{align}
\chi_{1}\left(  \alpha\right)   &  =\frac{e\left(  1+R\right)  }{2DR^{2}%
},\label{f72a}\\
\chi_{2}\left(  \alpha\right)   &  =\frac{e\left(  1-R\right)  }{2DR^{2}%
},\label{3.47}\\
\chi_{3}\left(  \alpha\right)   &  =\left(  1-R^{2}\right)  \frac{e}{DR^{2}},
\label{f72c}%
\end{align}
which give rise to the solutions
\begin{align}
\omega_{1}\left(  \alpha\right)   &  =n^{2}\eta^{\prime}\frac{e\left(
1+R\right)  }{2DR}\label{f73a}\\
\omega_{2}\left(  \alpha\right)   &  =-n^{2}\eta^{\prime}\frac{e\left(
1-R\right)  }{2DR},\label{f73b}\\
\psi_{1}\left(  \alpha\right)   &  =n^{2}\eta^{\prime}\frac{\sqrt{1-R^{2}}}%
{2}\frac{e\left(  1+R\right)  }{DR^{2}},\label{f74a}\\
\psi_{2}\left(  \alpha\right)   &  =n^{2}\eta^{\prime}\frac{\sqrt{1-R^{2}}}%
{2}\frac{e\left(  1-R\right)  }{DR^{2}},\label{f74b}\\
\psi_{3}\left(  \alpha\right)   &  =-n^{2}\eta^{\prime}\frac{2\alpha
\mu^{\prime}}{\kappa^{2}}\frac{e}{DR^{2}}=-n^{2}\eta^{\prime}\sqrt{1-R^{2}%
}\frac{e}{DR^{2}},\label{f74d}\\
\xi_{1}\left(  \alpha\right)   &  =\frac{\sqrt{1-R^{2}}}{2}\frac{e}{DR^{2}%
},\label{f75d}\\
\xi_{2}\left(  \alpha\right)   &  =\frac{\sqrt{1-R^{2}}}{2}\frac{e}{DR^{2}%
},\label{f75a}\\
\xi_{3}\left(  \alpha\right)   &  =-\sqrt{1-R^{2}}\frac{e}{DR^{2}%
},\label{f75b}\\
\xi_{4}  &  =0. \label{f75c}%
\end{align}
\bigskip

Finally, the solutions for $g\left(  \alpha,\rho\right)  $, etc. are obtained:%
\begin{align}
g\left(  \alpha,\rho\right)   &  =\frac{n^{2}e\eta^{\prime}}{D}\left[
\frac{\left(  1+R\right)  }{2R}K_{0}\left(  \lambda_{1}\rho\right)
-\frac{\left(  1-R\right)  }{2R}K_{0}\left(  \lambda_{2}\rho\right)  \right]
,\label{f76a}\\
u\left(  \alpha,\rho\right)   &  =\frac{n^{2}e\eta^{\prime}\mu^{\prime}%
}{D\kappa^{2}}\left[  \frac{\alpha\left(  1+R\right)  }{R^{2}}K_{0}\left(
\lambda_{1}\rho\right)  +\frac{\alpha\left(  1-R\right)  }{R^{2}}K_{0}\left(
\lambda_{2}\rho\right)  -\frac{2\alpha}{R^{2}}K_{0}\left(  \lambda_{3}%
\rho\right)  \right]  ,\label{f76b}\\
\gamma\left(  \alpha,\rho\right)   &  =\frac{e}{D}\left[  \frac{\left(
1+R\right)  }{2R^{2}}K_{0}\left(  \lambda_{1}\rho\right)  +\frac{\left(
1-R\right)  }{2R^{2}}K_{0}\left(  \lambda_{2}\rho\right)  -\frac{1-R^{2}%
}{R^{2}}K_{0}\left(  \lambda_{3}\rho\right)  \right]  ,\label{f76c}\\
y\left(  \alpha,\rho\right)   &  =\frac{e\mu^{\prime}}{D\kappa^{2}}\left[
\frac{\alpha}{R^{2}}K_{0}\left(  \lambda_{1}\rho\right)  +\frac{\alpha}{R^{2}%
}K_{0}\left(  \lambda_{2}\rho\right)  -\frac{2\alpha}{R^{2}}K_{0}\left(
\lambda_{3}\rho\right)  \right]  , \label{f76d}%
\end{align}
with which the Fourier transforms of the final solutions for the distribution
functions and potentials can be constructed in the manner of Eqs. (\ref{f43})
and (\ref{f44}). For this purpose it is convenient to define the following
abbreviations for the integrals:%
\begin{align}
G_{1}\left(  \rho\right)   &  =\frac{2}{\pi}\int_{0}^{\infty}d\alpha
\frac{\left(  1+R\right)  }{2R}\cos\left(  \alpha x\right)  K_{0}\left(
\lambda_{1}\rho\right)  ,\label{f77a}\\
G_{2}\left(  \rho\right)   &  =-\frac{2}{\pi}\int_{0}^{\infty}d\alpha
\frac{\left(  1-R\right)  }{2R}\cos\left(  \alpha x\right)  K_{0}\left(
\lambda_{2}\rho\right)  , \label{f77b}%
\end{align}%
\begin{align}
U_{1}\left(  \rho\right)   &  =\frac{2}{\pi}\int_{0}^{\infty}d\alpha
\frac{\alpha\left(  1+R\right)  }{R^{2}}\sin\left(  \alpha x\right)
K_{0}\left(  \lambda_{1}\rho\right)  ,\label{f78a}\\
U_{2}\left(  \rho\right)   &  =\frac{2}{\pi}\int_{0}^{\infty}d\alpha
\frac{\alpha\left(  1-R\right)  }{R^{2}}\sin\left(  \alpha x\right)
K_{0}\left(  \lambda_{2}\rho\right)  ,\label{f78b}\\
U_{3}\left(  \rho\right)   &  =-\frac{2}{\pi}\int_{0}^{\infty}d\alpha
\frac{2\alpha}{R^{2}}\sin\left(  \alpha x\right)  K_{0}\left(  \lambda_{3}%
\rho\right)  , \label{f78c}%
\end{align}%
\begin{align}
\Gamma_{1}\left(  \rho\right)   &  =\frac{2}{\pi}\int_{0}^{\infty}d\alpha
\frac{\left(  1+R\right)  }{2R^{2}}\cos\left(  \alpha x\right)  K_{0}\left(
\lambda_{1}\rho\right)  ,\label{f79}\\
\Gamma_{2}\left(  \rho\right)   &  =\frac{2}{\pi}\int_{0}^{\infty}d\alpha
\frac{\left(  1-R\right)  }{2R^{2}}\cos\left(  \alpha x\right)  K_{0}\left(
\lambda_{2}\rho\right)  ,\label{f79b}\\
\Gamma_{3}\left(  \rho\right)   &  =-\frac{2}{\pi}\int_{0}^{\infty}%
d\alpha\frac{1-R^{2}}{R^{2}}\cos\left(  \alpha x\right)  K_{0}\left(
\lambda_{3}\rho\right)  , \label{f79c}%
\end{align}%
\begin{align}
Y_{1}\left(  \rho\right)   &  =\frac{2}{\pi}\int_{0}^{\infty}d\alpha
\frac{\alpha}{R^{2}}\sin\left(  \alpha x\right)  K_{0}\left(  \lambda_{1}%
\rho\right)  ,\label{f80a}\\
Y_{2}\left(  \rho\right)   &  =\frac{2}{\pi}\int_{0}^{\infty}d\alpha
\frac{\alpha}{R^{2}}\sin\left(  \alpha x\right)  K_{0}\left(  \lambda_{2}%
\rho\right)  ,\label{f80b}\\
Y_{3}\left(  \rho\right)   &  =-\frac{2}{\pi}\int_{0}^{\infty}d\alpha
\frac{2\alpha}{R^{2}}\sin\left(  \alpha x\right)  K_{0}\left(  \lambda_{3}%
\rho\right)  . \label{f80c}%
\end{align}
The formal solutions for the distribution functions and potentials are given
by the forms:%
\begin{align}
f_{ii}\left(  \mathbf{r}\right)   &  =n^{2}+\frac{z_{i}e^{2}n^{2}}{Dk_{B}%
T}\left[  \Gamma_{1}\left(  \rho\right)  +\Gamma_{2}\left(  \rho\right)
+\Gamma_{3}\left(  \rho\right)  \right]  ,\label{f81a}\\
f_{jj}\left(  \mathbf{r}\right)   &  =n^{2}-\frac{z_{j}e^{2}n^{2}}{Dk_{B}%
T}\left[  \Gamma_{1}\left(  \rho\right)  +\Gamma_{2}\left(  \rho\right)
+\Gamma_{3}\left(  \rho\right)  \right]  ,\label{f81b}\\
f_{ij}\left(  \mp\mathbf{r}\right)   &  =f_{ji}\left(  \pm\mathbf{r}\right)
=\frac{n^{2}e\eta^{\prime}}{D}\left[  G_{1}\left(  \rho\right)  +G_{2}\left(
\rho\right)  \right] \nonumber\\
&  \qquad\qquad\qquad\pm\frac{n^{2}e\eta^{\prime}\mu^{\prime}}{D\kappa^{2}%
}\left[  U_{1}\left(  \rho\right)  +U_{2}\left(  \rho\right)  +U_{3}\left(
\rho\right)  \right]  , \label{f81c}%
\end{align}
and%
\begin{align}
\psi_{j}\left(  \mathbf{r}\right)   &  =-\psi_{i}\left(  -\mathbf{r}\right)
\nonumber\\
&  =\frac{e}{D}\left[  \Gamma_{1}\left(  \rho\right)  +\Gamma_{2}\left(
\rho\right)  +\Gamma_{3}\left(  \rho\right)  \right]  +\frac{e\mu^{\prime}%
}{D\kappa^{2}}\left[  Y_{1}\left(  \rho\right)  +Y_{2}\left(  \rho\right)
+Y_{3}\left(  \rho\right)  \right]  ,\label{f82a}\\
\psi_{j}\left(  -\mathbf{r}\right)   &  =-\psi_{i}\left(  \mathbf{r}\right)
\nonumber\\
&  =\frac{e}{D}\left[  \Gamma_{1}\left(  \rho\right)  +\Gamma_{2}\left(
\rho\right)  +\Gamma_{3}\left(  \rho\right)  \right]  -\frac{e\mu^{\prime}%
}{D\kappa^{2}}\left[  Y_{1}\left(  \rho\right)  +Y_{2}\left(  \rho\right)
+Y_{3}\left(  \rho\right)  \right]  . \label{f82b}%
\end{align}
These formal solutions given in terms of quadratures of the zeroth-order
Bessel functions of second kind can be used to calculate various local mean
quantities associated with transport in binary electrolyte solutions. Since
the Bessel functions are not a trivial function to compute the quadratures
with, it is useful to review briefly their mathematical properties we can make
use of to evaluate the quadratures in question.

The Bessel function $K_{\nu}\left(  z\right)  $ obeys the differential
equation\cite{watson,abramowitz}%
\begin{equation}
\left[  \frac{1}{z}\frac{d}{dz}z\frac{d}{dz}-\left(  1+\frac{\nu^{2}}{z^{2}%
}\right)  \right]  K_{\nu}\left(  z\right)  =0. \label{a1}%
\end{equation}
It is a regular function in complex $z$ plane cut from $z=0$ to $z=-\infty$.
One of its integral representation is given by%
\begin{equation}
K_{\nu}\left(  z\right)  =\int_{0}^{\infty}dte^{-z\cosh t}\cosh\left(  \nu
t\right)  \qquad\left(  \left\vert \arg z\right\vert <\frac{\pi}{2}\right)  .
\label{a2}%
\end{equation}
We will make use of this integral representation. We also note one of the
recurrence relations
\begin{equation}
K_{\nu}^{\prime}\left(  z\right)  =-K_{\nu+1}\left(  z\right)  +\frac{\nu}%
{z}K_{\nu}\left(  z\right)  . \label{a3}%
\end{equation}
Here the prime denotes differentiation with respect to $z$. This recurrence
relation implies%
\begin{equation}
K_{0}^{\prime}\left(  z\right)  =-K_{1}\left(  z\right)  . \label{a4}%
\end{equation}
In particular, $K_{0}\left(  z\right)  $ may be represented by the ascending
series%
\begin{equation}
K_{0}\left(  z\right)  =-\left[  \ln\left(  \frac{z}{2}\right)  +\gamma
\right]  I_{0}(z)+\sum_{i=1}^{\infty}\left(  1+\frac{1}{2}+\cdots+\frac{1}%
{i}\right)  \frac{\left(  \frac{z}{2}\right)  ^{2n}}{\left(  i!\right)  ^{2}},
\label{a5}%
\end{equation}
where $I_{0}(z)$ is the other independent solution of Eq. (\ref{a1}), which is
given by the ascending series%
\begin{equation}
I_{0}(z)=\sum_{i=0}^{\infty}\frac{\left(  \frac{z}{2}\right)  ^{2n}}{\left(
i!\right)  ^{2}}. \label{a6}%
\end{equation}
This function is an entire function if the cut $z$ plane. The asymptotic
behaviors of $K_{0}\left(  z\right)  $ and $I_{0}(z)$ are follows: as
$\left\vert z\right\vert \rightarrow\infty$%
\begin{align}
K_{0}\left(  z\right)   &  =\sqrt{\frac{\pi}{2z}}e^{-z}\left[  1+O(z^{-1}%
)\right]  ,\label{a7a}\\
I_{0}\left(  z\right)   &  =\sqrt{\frac{1}{2\pi z}}e^{z}\left[  1+O(z^{-1}%
)\right]  \label{a7b}%
\end{align}
for $\left\vert \arg z\right\vert <\frac{\pi}{2}$. It is useful to note that
the presence of $\ln\left(  \frac{z}{2}\right)  $ in $K_{0}\left(  z\right)  $
clearly indicates that $K_{0}\left(  z\right)  $ is not only discontinuous
across the branch cut, but also diverges logarithmically as $z\rightarrow0$.
If $K_{0}\left(  z\right)  $ is analytically continued across the cut, we
obtain, for example,
\begin{equation}
K_{0}\left(  e^{i\pi}z\right)  =K_{0}\left(  z\right)  -\pi iI_{0}\left(
z\right)  . \label{a8}%
\end{equation}
This relation is easily seen to be true from Eq. (\ref{a5}).

Wilson makes use of only Eqs. (\ref{a4}) and (\ref{a5}) to evaluate integrals
for the velocity and ionic field, which involve the Bessel functions. This
analytically continued form as well as the properties presented above can be
made use of for exact evaluation of the quadratures presented in Eqs.
(\ref{f77a})--(\ref{f80c}), as will be shown elsewhere\cite{euwien}.

\section{\textbf{Electrophoresis}}

Onsager in his 1927 paper\cite{onsager} proposed that there exist two effects
in electrolytic conduction, one the electrophoretic effect and the other the
relaxation time effect. In Ref. \cite{onsager}, the electrophoretic effect was
treated by making use of the Stokes law\cite{landau} and the relaxation time
effect by estimating relaxation time by a heuristic argument\cite{harned}
first used by J. C. Maxwell. In the Onsager--Wilson theory the electrophoresis
effect is calculated by solving the Navier--Stokes equations\cite{landau} and
the time of relaxation effect by calculating the ionic field using the
solutions presented in the previous section. In this section, we discuss the
electrophoresis effect.

The Navier--Stokes equations\cite{landau} necessary for the calculation of the
effect may be cast into the forms%
\begin{align}
\eta_{0}\mathbf{\nabla\times\nabla}\times\mathbf{v\,}  &  \mathbf{=\,}%
\mathbf{-\nabla}p+\mathbf{F,}\label{e1}\\
\mathbf{\nabla\cdot v}  &  =0, \label{e2}%
\end{align}
where $\mathbf{F}$ is the local external force, $p$ is the pressure, and
$\eta_{0}$ is the shear viscosity of the fluid. In solving these equations
Wilson makes the following assumption: \textit{the forces exerted by the field
due to an ion and the ion atmosphere upon the charges in the ion atmosphere
may be neglected, because they are proportional to the square of the charges
of the central ion.} This means the external forces include forces linearly
proportional to the external force $X$ and the ionic potentials. Therefore
\[
\mathbf{F=}\rho\mathbf{X}%
\]
and since the charge density $j$ is described by the Poisson equation
\begin{equation}
F_{x}=-\frac{DX}{4\pi}\nabla^{2}\psi_{j}\left(  \mathbf{r}\right)  \label{e2f}%
\end{equation}
and the Navier--Stokes equation may be put in the form%
\begin{equation}
\eta_{0}\mathbf{\nabla\times\nabla}\times\mathbf{v=-\nabla}p+\mathbf{F}
\label{e3}%
\end{equation}
for a binary electrolyte.

To solve Eq. (\ref{e3}) we look for a vector $\mathbf{a}$ satisfying
\begin{equation}
\eta_{0}\nabla^{2}\left(  \nabla^{2}\mathbf{a}\right)  =\mathbf{F,} \label{e4}%
\end{equation}
where $\mathbf{F}$ is the vector form of Eq. (\ref{e2f}). Eliminating
$\mathbf{F}$ between Eq. (\ref{e1}) and Eq. (\ref{e4}), we obtain%
\begin{equation}
\eta_{0}\mathbf{\nabla\times\nabla}\times\mathbf{v+\nabla}p-\eta_{0}\nabla
^{2}\left(  \nabla^{2}\mathbf{a}\right)  =0. \label{e5}%
\end{equation}
Noting the identity in vector algebra\cite{margenau}%
\[
\nabla\times\nabla\times\left(  \nabla^{2}\mathbf{a}\right)  =\nabla\left[
\nabla^{2}\left(  \nabla\cdot\mathbf{a}\right)  \right]  -\nabla^{2}\left(
\nabla^{2}\mathbf{a}\right)  ,
\]
it is possible to express Eq. (\ref{e5}) as%
\begin{equation}
\eta_{0}\nabla\times\nabla\left(  \mathbf{v+}\nabla^{2}\mathbf{a}\right)
\mathbf{+\nabla}\left[  p-\eta_{0}\nabla^{2}\left(  \nabla\cdot\mathbf{a}%
\right)  \right]  =0, \label{e6}%
\end{equation}
but%
\begin{equation}
\nabla\times\nabla\times\left(  \nabla^{2}\mathbf{a}\right)  =\nabla
\times\nabla\times\left[  \nabla^{2}\mathbf{a-}\nabla\left(  \nabla
\cdot\mathbf{a}\right)  \right]  =-\nabla\times\nabla\times\nabla\times
\nabla\times\mathbf{a} \label{e6a}%
\end{equation}
because%
\[
\nabla\times\nabla\times\nabla\left(  \nabla\cdot\mathbf{a}\right)  =0.
\]
Inserting Eq. (\ref{e6a}) into Eq. (\ref{e6}), we obtain%
\begin{equation}
\eta_{0}\nabla\times\nabla\times\left(  \mathbf{v-}\nabla\times\nabla
\times\mathbf{a}\right)  +\mathbf{\nabla}\left[  p-\eta_{0}\nabla^{2}\left(
\nabla\cdot\mathbf{a}\right)  \right]  =0. \label{e6b}%
\end{equation}
Since two terms in Eq. (\ref{e6}) are independent, it is found%
\begin{align}
\mathbf{v\,}  &  \mathbf{=\,}\nabla\times\nabla\times\mathbf{a,}\label{e7}\\
p  &  =p_{0}+\eta_{0}\nabla^{2}\left(  \nabla\cdot\mathbf{a}\right)  .
\label{e8}%
\end{align}
In Eq. (\ref{e7}) for $\mathbf{v}$, the sign error in Wilson's thesis is
corrected, and $\eta_{0}$ is also added in Eq. (\ref{e8}). In any case, the
solution of Eq. (\ref{e4}) provides the velocity field and the local pressure
arising from the local potentials. It, however, should be noted that the
solution of the equation%
\[
\mathbf{\nabla}\left[  p-\eta_{0}\nabla^{2}\left(  \nabla\cdot\mathbf{a}%
\right)  \right]  =0
\]
is more generally%
\[
p-\eta_{0}\nabla^{2}\left(  \nabla\cdot\mathbf{a}\right)  =p_{0},
\]
where $p_{0}$ must be uniform in space in the hydrodynamic scale for the
Navier--Stokes equation,\ so that%
\[
\nabla p_{0}=0.
\]
Therefore, $p_{0}$ may be regarded as a uniform equilibrium pressure in the
absence of the external field. Thus, to be correct, $p_{0}$ is added to Eq.
(\ref{e8}). The $p_{0}$ is missing in Wilson's expression (\ref{e8}) for $p$.
Here $p_{0}$ is the equilibrium pressure of the solution that may be given,
for example, by the virial form of pressure\cite{hill}:%
\begin{equation}
p_{0}=nk_{B}T-\frac{2\pi}{3}\sum_{i<j}n_{i}n_{j}\int_{0}^{\infty}drr^{3}%
\frac{du_{ij}}{dr}g_{ij}^{0}\left(  r\right)  , \label{e8b}%
\end{equation}
where $u_{ij}$ is the intermolecular potential of pair $\left(  i,j\right)  $
and $g_{ij}^{0}\left(  r\right)  $ is the equilibrium pair correlation
function. The index $i$ runs over the species in the fluid, including ions.
Determination of $g_{ij}^{0}\left(  r\right)  $ should be made by following
the modern theory of equilibrium Coulomb (ionic) fluids.\cite{coulombfluid}

It is now necessary to find the vector $\mathbf{a}$ by solving Eq. (\ref{e4}).
Since the local force $F_{x}$ is a linear combination of the solutions for
potentials, it may be written in the form%
\begin{equation}
F_{x}\left(  x,\rho\right)  =\eta_{0}\sum_{i=1}^{3}\mathbf{C}_{i}K_{0}\left(
\lambda_{i}\rho\right)  +\eta_{0}\sum_{i=1}^{3}\mathbf{S}_{i}K_{0}\left(
\lambda_{i}\rho\right)  , \label{e9}%
\end{equation}
where $\mathbf{C}_{i}$ and $\mathbf{S}_{i}$ are integral operators defined in
$\alpha$ space as below:%
\begin{align}
\mathbf{C}_{1}  &  =-\frac{eX}{2\pi^{2}\eta_{0}}\int_{0}^{\infty}d\alpha
\cos\left(  \alpha x\right)  \frac{\left(  1+R\right)  \left(  \lambda_{1}%
^{2}-\alpha^{2}\right)  }{2R^{2}},\nonumber\\
\mathbf{C}_{2}  &  =-\frac{eX}{2\pi^{2}\eta_{0}}\int_{0}^{\infty}d\alpha
\cos\left(  \alpha x\right)  \frac{\left(  1-R\right)  \left(  \lambda_{2}%
^{2}-\alpha^{2}\right)  }{2R^{2}},\label{e10a}\\
\mathbf{C}_{3}  &  =\frac{eX}{2\pi^{2}\eta_{0}}\int_{0}^{\infty}d\alpha
\cos\left(  \alpha x\right)  \frac{\left(  1-R^{2}\right)  \left(  \lambda
_{3}^{2}-\alpha^{2}\right)  }{R^{2}},\nonumber\\
\mathbf{S}_{1}  &  =-\frac{eX\mu^{\prime}}{2\pi^{2}\eta_{0}\kappa^{2}}\int
_{0}^{\infty}d\alpha\sin\left(  \alpha x\right)  \frac{\alpha\left(
\lambda_{1}^{2}-\alpha^{2}\right)  }{R^{2}},\nonumber\\
\mathbf{S}_{2}  &  =-\frac{eX\mu^{\prime}}{2\pi^{2}\eta_{0}\kappa^{2}}\int
_{0}^{\infty}d\alpha\sin\left(  \alpha x\right)  \frac{\alpha\left(
\lambda_{2}^{2}-\alpha^{2}\right)  }{R^{2}},\label{e10b}\\
\mathbf{S}_{3}  &  =\frac{eX\mu^{\prime}}{2\pi^{2}\eta_{0}\kappa^{2}}\int
_{0}^{\infty}d\alpha\sin\left(  \alpha x\right)  \frac{2\alpha\left(
\lambda_{3}^{2}-\alpha^{2}\right)  }{R^{2}}.\nonumber
\end{align}
With $F_{x}$ so expressed, the equation for the $x\mathbf{\ }$component of
vector $\mathbf{a}$, Eq. (\ref{e4}), reads%
\begin{equation}
\nabla^{2}\left(  \nabla^{2}a_{x}\right)  =\sum_{i=1}^{3}\mathbf{C}_{i}%
K_{0}\left(  \lambda_{i}\rho\right)  +\sum_{i=1}^{3}\mathbf{S}_{i}K_{0}\left(
\lambda_{i}\rho\right)  . \label{e11}%
\end{equation}
This is solved by%
\begin{equation}
\nabla^{2}a_{x}=\sum_{i=1}^{3}\left(  \mathbf{C}_{i}+\mathbf{S}_{i}\right)
\frac{K_{0}\left(  \lambda_{i}\rho\right)  }{\lambda_{i}^{2}-\alpha^{2}%
}+A^{\ast}, \label{e12}%
\end{equation}
where $A^{\ast}$ is the solution of the homogeneous equation%
\begin{equation}
\nabla^{2}\left(  \nabla^{2}A^{\ast}\right)  =0. \label{e13}%
\end{equation}
It must be also a linear combination of $K_{0}\left(  \lambda_{i}\rho\right)
$, but since it must satisfy the boundary conditions we choose it in the form%
\begin{equation}
A^{\ast}=-\sum_{i=1}^{3}\left(  \mathbf{C}_{i}+\mathbf{S}_{i}\right)
\frac{K_{0}\left(  \alpha\rho\right)  }{\lambda_{i}^{2}-\alpha^{2}}.
\label{e14}%
\end{equation}
Therefore we have%
\begin{equation}
\nabla^{2}a_{x}=\sum_{i=1}^{3}\left(  \mathbf{C}_{i}+\mathbf{S}_{i}\right)
\frac{\left[  K_{0}\left(  \lambda_{i}\rho\right)  -K_{0}\left(  \alpha
\rho\right)  \right]  }{\lambda_{i}^{2}-\alpha^{2}}. \label{e15}%
\end{equation}
It should be remarked that this solution is opposite in sign to Wilson's. The
solution of this equation must be also a linear combination of the Bessel
functions $K_{0}\left(  \lambda_{i}\rho\right)  $ and $K_{0}(\alpha\rho)$.
Wilson does not make the solution process for this equation explicit. Here we
make it transparent. The solution is sought in the form%
\begin{equation}
a_{x}=\sum_{i=1}^{3}\left(  \mathbf{C}_{i}+\mathbf{S}_{i}\right)  \frac
{1}{\lambda_{i}^{2}-\alpha^{2}}\left\{  b_{1}\left[  K_{0}\left(  \lambda
_{i}\rho\right)  -K_{0}\left(  \alpha\rho\right)  \right]  +b_{2}\left[
K_{0}\left(  \beta_{i}\rho\right)  -K_{0}\left(  \alpha\rho\right)  \right]
\right\}  , \label{e16}%
\end{equation}
where $\beta$ is a parameter to be determined such that the right hand side is
a solution of Eq. (\ref{e15}). On inserting this into Eq. (\ref{e15}) we find%
\[
\left[  b_{1}\left(  \lambda_{i}^{2}-\alpha^{2}\right)  -1\right]
K_{0}(\lambda_{i}\rho)+b_{2}\left(  \beta_{i}^{2}-\alpha^{2}\right)
K_{0}(\beta_{i}\rho)+K_{0}(\alpha\rho)=0.
\]
Now, $b_{1}$ and $b_{2}$ are chosen such that%
\begin{equation}
b_{1}=\frac{1}{\lambda_{i}^{2}-\alpha^{2}} \label{e17a}%
\end{equation}
and%
\[
\lim_{\beta_{i}\rightarrow\alpha}\left[  b_{2}\left(  \beta_{i}^{2}-\alpha
^{2}\right)  K_{0}(\beta_{i}\rho)+K_{0}(\alpha\rho)\right]  =0,
\]
that is,%
\begin{equation}
\lim_{\beta_{i}\rightarrow\alpha}b_{2}\left(  \beta_{i}^{2}-\alpha^{2}\right)
K_{0}(\beta_{i}\rho)=-K_{0}(\alpha\rho). \label{e17b}%
\end{equation}
Then we find%
\begin{equation}
b_{2}=-\frac{1}{\left(  \beta_{i}^{2}-\alpha^{2}\right)  }. \label{e17c}%
\end{equation}
Now, since
\begin{align}
\lim_{\beta_{i}\rightarrow\alpha}\frac{1}{\left(  \beta_{i}^{2}-\alpha
^{2}\right)  }\left[  K_{0}\left(  \beta_{i}\rho\right)  -K_{0}\left(
\alpha\rho\right)  \right]   &  =\frac{1}{2\alpha}\lim_{\beta_{i}%
\rightarrow\alpha}\frac{\left[  K_{0}\left(  \beta_{i}\rho\right)
-K_{0}\left(  \alpha\rho\right)  \right]  }{\left(  \beta_{i}+\alpha\right)
}\nonumber\\
&  =\frac{\rho}{2\alpha}K_{0}^{\prime}\left(  \alpha\rho\right) \nonumber\\
&  =-\frac{\rho}{2\alpha}K_{1}\left(  \alpha\rho\right)  \label{e18}%
\end{align}
in which we have used recurrence relation (\ref{a4}) for the last equality and
the prime denotes the derivative with respect to the argument $\alpha\rho$, we
finally obtain the solution of Eq. (\ref{e15}) in the form
\begin{equation}
a_{x}=\sum_{i=1}^{3}\left(  \mathbf{C}_{i}+\mathbf{S}_{i}\right)  \left\{
\frac{\left[  K_{0}\left(  \lambda_{i}\rho\right)  -K_{0}\left(  \alpha
\rho\right)  \right]  }{\left(  \lambda_{i}^{2}-\alpha^{2}\right)  ^{2}}%
+\frac{\rho K_{1}\left(  \alpha\rho\right)  }{2\alpha\left(  \lambda_{i}%
^{2}-\alpha^{2}\right)  }\right\}  . \label{e19}%
\end{equation}
With this result for $a_{x}$ it is possible to calculate the velocity by using
the expressions%
\begin{align}
\operatorname{div}\mathbf{a}  &  \mathbf{=}\sum_{i=1}^{3}\left[
\mathbf{-}C_{i}+S_{i}\right]  \left\{  \frac{\alpha}{\left(  \lambda_{i}%
^{2}-\alpha^{2}\right)  ^{2}}\left[  K_{0}(\lambda_{i}\rho)-K_{0}(\alpha
\rho)\right]  +\frac{\rho}{2\left(  \lambda_{i}^{2}-\alpha^{2}\right)  }%
K_{1}(\alpha\rho)\right\}  ,\label{e20}\\
\nabla_{x}\operatorname{div}\mathbf{a}  &  \mathbf{=-}\sum_{i=1}^{3}\left[
C_{i}+S_{i}\right]  \left\{  \frac{\alpha^{2}}{\left(  \lambda_{i}^{2}%
-\alpha^{2}\right)  ^{2}}\left[  K_{0}(\lambda_{i}\rho)-K_{0}(\alpha
\rho)\right]  +\frac{\alpha\rho}{2\left(  \lambda_{i}^{2}-\alpha^{2}\right)
}K_{1}(\alpha\rho)\right\}  . \label{e21}%
\end{align}

\subsection{\textbf{Axial Velocity}}

Inserting the formula for $a_{x}$ into Eq. (\ref{e7}) and Eq. (\ref{e8}) and
making use of Eqs. (\ref{e20}) and (\ref{e21}) we readily obtain the velocity
field. The axial velocity is finally obtained as below:\
\begin{align}
\mathbf{v}_{x}\left(  x,\rho,0\right)   &  =\frac{eX}{2\pi^{2}\eta_{0}}%
\int_{0}^{\infty}d\alpha\cos\left(  \alpha x\right)  \times\nonumber\\
&  \quad\left\{  \frac{\left(  1+R\right)  \lambda_{1}^{2}}{2R^{2}\left(
\lambda_{1}^{2}-\alpha^{2}\right)  }\left[  K_{0}(\lambda_{1}\rho
)-K_{0}(\alpha\rho)\right]  \right. \nonumber\\
&  +\frac{\left(  1-R\right)  \lambda_{2}^{2}}{2R^{2}\left(  \lambda_{2}%
^{2}-\alpha^{2}\right)  }\left[  K_{0}(\lambda_{2}\rho)-K_{0}(\alpha
\rho)\right] \nonumber\\
&  \left.  -\frac{\left(  1-R^{2}\right)  \lambda_{3}^{2}}{R^{2}\left(
\lambda_{3}^{2}-\alpha^{2}\right)  }\left[  K_{0}(\lambda_{3}\rho
)-K_{0}(\alpha\rho)\right]  +\frac{\alpha\rho}{2}K_{1}(\alpha\rho)\right\}
\nonumber\\
&  \mathbf{-}\frac{eX\mu^{\prime}}{2\pi^{2}\eta_{0}\kappa^{2}}\int_{0}%
^{\infty}d\alpha\sin\left(  \alpha x\right)  \left\{  \frac{\alpha\lambda
_{1}^{2}}{R^{2}\left(  \lambda_{1}^{2}-\alpha^{2}\right)  }\left[
K_{0}(\lambda_{1}\rho)-K_{0}(\alpha\rho)\right]  \right. \nonumber\\
&  +\frac{\alpha\lambda_{2}^{2}}{R^{2}\left(  \lambda_{2}^{2}-\alpha
^{2}\right)  }\left[  K_{0}(\lambda_{2}\rho)-K_{0}(\alpha\rho)\right]
\nonumber\\
&  \left.  -\frac{2\alpha\lambda_{3}^{2}}{R^{2}\left(  \lambda_{3}^{2}%
-\alpha^{2}\right)  }\left[  K_{0}(\lambda_{3}\rho)-K_{0}(\alpha\rho)\right]
\right\}  . \label{e22}%
\end{align}
in which\ the integral operators $\mathbf{C}_{i}$ and $\mathbf{S}_{i}$ defined
in Eqs. (\ref{e10a}) and (\ref{e10b}) are made explicit. Further simplifying
it, we obtain%
\begin{align}
\mathbf{v}_{x}\left(  x,\rho,0\right)   &  =\frac{eX}{2\pi^{2}\eta_{0}}%
\int_{0}^{\infty}d\alpha\cos\left(  \alpha x\right)  \times\nonumber\\
&  \qquad\qquad\left\{  \frac{1}{\kappa^{2}R^{2}}\left[  \lambda_{1}^{2}%
K_{0}(\lambda_{1}\rho)+\lambda_{2}^{2}K_{0}(\lambda_{2}\rho)-2\lambda_{3}%
^{2}\left(  1-R^{2}\right)  K_{0}(\lambda_{3}\rho)\right]  \right. \nonumber\\
&  \qquad\qquad\left.  -\frac{2\lambda_{3}^{2}}{\kappa^{2}}K_{0}(\alpha
\rho)+\frac{\alpha\rho}{2}K_{1}(\alpha\rho)\right\} \nonumber\\
&  \mathbf{-}\frac{eX\mu^{\prime}}{2\pi^{2}\eta_{0}\kappa^{2}}\int_{0}%
^{\infty}d\alpha\sin\left(  \alpha x\right)  \frac{2\alpha}{\kappa^{2}R^{2}%
}\left[  \lambda_{1}^{2}K_{0}(\lambda_{1}\rho)+\lambda_{2}^{2}K_{0}%
(\lambda_{2}\rho)-2\lambda_{3}^{2}K_{0}(\lambda_{3}\rho)\right]  .
\label{e22a}%
\end{align}

Since the quantity of interest for electrophoresis is the axial velocity at
the center ion (the coordinate origin) Wilson at this point consider the
special case of $\mathbf{v}_{x}\left(  x,\rho,0\right)  $ at $x=0$ and
$\rho=0$ before fully evaluating the integrals in Eq. (\ref{e22}). That is, if
$x=0$ and $\rho=0$ are taken, the axial velocity component at the center ion
is given by%
\begin{align}
v_{x}\left(  0,0,0\right)   &  =\frac{eX}{2\pi^{2}\eta_{0}}\int_{0}^{\infty
}d\alpha\left\{  -\frac{\lambda_{1}^{2}\left(  1+R\right)  }{2R^{2}\left(
\lambda_{1}^{2}-\alpha^{2}\right)  }\ln\left(  \frac{\lambda_{1}}{\alpha
}\right)  \right. \nonumber\\
&  \left.  -\frac{\lambda_{2}^{2}\left(  1-R\right)  }{2R^{2}\left(
\lambda_{2}^{2}-\alpha^{2}\right)  }\ln\left(  \frac{\lambda_{2}}{\alpha
}\right)  +\frac{\lambda_{3}^{2}\left(  1-R^{2}\right)  }{R^{2}\left(
\lambda_{3}^{2}-\alpha^{2}\right)  }\ln\left(  \frac{\lambda_{3}}{\alpha
}\right)  -\frac{1}{2}\right\}  . \label{e23}%
\end{align}
He then evaluates these integrals by using the method of contour in complex
$\alpha$ plane. For the purpose it should be noted that
\begin{align*}
\frac{\lambda_{1}^{2}\left(  1+R\right)  }{2R^{2}\left(  \lambda_{1}%
^{2}-\alpha^{2}\right)  }  &  =\frac{\kappa^{2}}{2}\frac{1+t^{2}%
+\sqrt{1-2x^{2}t^{2}}}{1-2x^{2}t^{2}},\\
\frac{\lambda_{2}^{2}\left(  1-R\right)  }{2R^{2}\left(  \lambda_{2}%
^{2}-\alpha^{2}\right)  }  &  =\frac{\kappa^{2}}{2}\frac{1+t^{2}%
-\sqrt{1-2x^{2}t^{2}}}{1-2x^{2}t^{2}},\\
\frac{\lambda_{3}^{2}\left(  1-R^{2}\right)  }{R^{2}\left(  \lambda_{3}%
^{2}-\alpha^{2}\right)  }  &  =x^{2}\frac{2t^{2}\left(  1+t^{2}\right)
}{1-2x^{2}t^{2}},
\end{align*}
where%
\[
x=\frac{\mu^{\prime}}{\kappa},\quad t=\frac{\sqrt{2}\alpha}{\kappa}.
\]
These expressions mean that there are simple poles at $t=\pm1/\left(  \sqrt
{2}x\right)  $ in the integrands. The arguments of the logarithmic functions
have branch points%
\begin{align*}
t  &  =\pm i\sqrt{2\left(  1+x^{2}\right)  }\quad\text{for }\lambda_{1}\\
t  &  =0,-\infty\quad\text{for }\lambda_{2}\\
t  &  =\pm i\quad\text{for }\lambda_{3}.
\end{align*}
Furthermore, the integrands of the first three integrals vanish to zero as
$\left\vert t\right\vert \rightarrow\infty$. Therefore the first three
integrals can be evaluated by applying the method of contour integration.
Wilson asserts that the integral of $-\frac{1}{2}$ for the last term can be
replaced by a contour integral $\mathcal{C}$ used for the other integrals and
the contour integral vanishes because the integrand is constant and hence%
\[
\int_{\mathcal{C}}dz\frac{1}{2}=0.
\]
Thus the last integral does not contribute to the electrophoretic effect. On
this ground, he obtains from Eq. (\ref{e23}) the formula for the
electrophoretic effect as represented by the formula%
\begin{equation}
v_{x}\left(  0,0,0\right)  =-\frac{Xe\kappa}{6\sqrt{2}\pi\eta_{0}}f(x)
\label{e24}%
\end{equation}
where%
\begin{align}
f(x)  &  =1+\frac{3}{4\sqrt{2}x}\left\{  2x^{2}\sinh^{-1}x+\sqrt{2}%
x-x\sqrt{1+x^{2}}\right. \nonumber\\
&  \left.  -\left(  1+2x^{2}\right)  \tan^{-1}\left(  \sqrt{2}x\right)
+\left(  1+2x^{2}\right)  \tan^{-1}\frac{x}{\sqrt{1+x^{2}}}\right\}  .
\label{e25}%
\end{align}
In effect, this result is correct, provided the last integral does not
contribute to the electrophoretic effect or negligible. However, as it stands,
Eq. (\ref{e23}) definitely contains a divergent integral.

In fact, in the case of $x=0$ the integral in Eq. (\ref{e22})
\[
-\int_{0}^{\infty}d\alpha\frac{\alpha\rho}{2}K_{1}(\alpha\rho)
\]
can be exactly evaluated for all values of $\rho>0$. By using $K_{0}^{\prime
}(z)=-K_{1}(z)$ and the integral representation for $K_{0}(\alpha\rho)$ it is
easy to show that%
\begin{equation}
\int_{0}^{\infty}d\alpha\alpha K_{1}(\alpha\rho)=\frac{\pi}{2\rho^{2}},
\label{e26a}%
\end{equation}
which indicates the integral is singular at $\rho=0$. Therefore $\rho=0$
should not be taken within the integral. In fact, if the ions are assumed to
be hard spheres as is usually assumed in the treatment of electrolytes---in
fact, this is what is done in the Debye--H\"{u}ckel theory underlying the
Onsager theory of conductivity---it may be appropriate to take the mean hard
sphere radius, say, $\sigma$ for the minimum value of $\rho$; $\rho=\sigma$ or
even $\rho=\kappa^{-1}$, the radius of the ion atmosphere. Then the axial
velocity emerges finite for $\sigma\leq\rho<\infty$. But then, it is necessary
to evaluate, for example, the integrals for $v_{x}\left(  0,\sigma,0\right)  $
with the form
\begin{align}
\mathbf{v}_{x}\left(  0,\sigma,0\right)   &  =\frac{eX}{2\pi^{2}\eta_{0}}%
\int_{0}^{\infty}d\alpha\left\{  \frac{\left(  1+R\right)  \lambda_{1}^{2}%
}{2R^{2}\left(  \lambda_{1}^{2}-\alpha^{2}\right)  }\left[  K_{0}(\lambda
_{1}\sigma)-K_{0}(\alpha\sigma)\right]  \right. \nonumber\\
&  +\frac{\left(  1-R\right)  \lambda_{2}^{2}}{2R^{2}\left(  \lambda_{2}%
^{2}-\alpha^{2}\right)  }\left[  K_{0}(\lambda_{2}\sigma)-K_{0}(\alpha
\sigma)\right] \nonumber\\
&  \left.  -\frac{\left(  1-R^{2}\right)  \lambda_{3}^{2}}{R^{2}\left(
\lambda_{3}^{2}-\alpha^{2}\right)  }\left[  K_{0}(\lambda_{3}\sigma
)-K_{0}(\alpha\sigma)\right]  \right\}  -\frac{3eX}{8\pi\eta_{0}\sigma}.
\label{e26}%
\end{align}
The integrals in this expression, however, cannot be evaluated by the method
used by Wilson. It would be straightforward to evaluate the integrals
numerically on computer. There then would be an extra contribution to the
electrophoretic effect arising from the last term on the right in Eq.
(\ref{e26}), and as a matter of fact, the terms associated with $K_{0}%
(\alpha\sigma)$ in Eq. (\ref{e26}) all together also contribute
\[
-\frac{eX\kappa}{8\pi\eta_{0}\left(  \kappa\rho\right)  ^{3}}%
\]
to the velocity. It also diverges as $\rho\rightarrow0$. If these are
evaluated at $\rho=\sigma$ or $\rho=\kappa^{-1}$, they are not negligible at
all and hence would modify Wilson's electrophoretic effect significantly.

It is therefore possible to conclude that Wilson's result for the
electrophoretic effect represents only the nondivergent part with the
divergent contribution arbitrarily discarded. This divergence difficulty might
have been the underlying reason for not publishing the one-dimensional Fourier
transform approach taken in Wilson's treatment in favor of a full
three-dimensional Fourier transform approach that had to wait twenty years
until the Onsager-Kim theory\cite{kim}; see page 146, section 8, Chapter 4 of
Ref. \cite{harned}.

\section{\textbf{Ionic Field}}

To calculate diffusion of ions it is necessary to know the ionic field of ion
$j$ located at $x=0$---the center of ion atmosphere. More precisely, we need
the contribution to the ionic field arising from the ion atmosphere. It is
given by%
\begin{equation}
\Delta\mathbf{X}\left(  \mathbf{r}\right)  \mathbf{=-\nabla}\psi_{j}\left(
\mathbf{r}\right)  . \label{e56}%
\end{equation}
Since $\psi_{j}\left(  \mathbf{r}\right)  $ is already known it is trivial to
compute. Inserting the formula for $\psi_{j}\left(  \mathbf{r}\right)  $
obtained earlier we find it in the form
\begin{align}
\Delta\mathbf{X}\left(  \mathbf{r}\right)   &  \mathbf{=}-\frac{2e}{\pi
D}\mathbf{\nabla}\int_{0}^{\infty}d\alpha\cos\left(  \alpha x\right)  \left[
\frac{\left(  1+R\right)  }{2R^{2}}K_{0}\left(  \lambda_{1}\rho\right)
+\frac{\left(  1-R\right)  }{2R^{2}}K_{0}\left(  \lambda_{2}\rho\right)
\right. \nonumber\\
&  \qquad\qquad\qquad\qquad\qquad\left.  -\frac{1-R^{2}}{R^{2}}K_{0}\left(
\lambda_{3}\rho\right)  \right] \nonumber\\
&  \qquad\mp\mathbf{\nabla}\frac{2e\mu^{\prime}}{\pi D\kappa^{2}}\int
_{0}^{\infty}d\alpha\sin\left(  \alpha x\right)  \left[  \frac{\alpha}{R^{2}%
}K_{0}\left(  \lambda_{1}\rho\right)  +\frac{\alpha}{R^{2}}K_{0}\left(
\lambda_{2}\rho\right)  \right. \label{e57}\\
&  \qquad\qquad\qquad\qquad\qquad\left.  -\frac{2\alpha}{R^{2}}K_{0}\left(
\lambda_{3}\rho\right)  \right] \nonumber
\end{align}
and hence%
\begin{align}
\Delta\mathbf{X}\left(  \mathbf{r}\right)   &  =\mathbf{e}_{x}\frac{2e}{\pi
D}\int_{0}^{\infty}d\alpha\alpha\sin\left(  \alpha x\right)  \left[
\frac{\left(  1+R\right)  }{2R^{2}}K_{0}\left(  \lambda_{1}\rho\right)
\right. \nonumber\\
&  \qquad\qquad\qquad\qquad\left.  +\frac{\left(  1-R\right)  }{2R^{2}}%
K_{0}\left(  \lambda_{2}\rho\right)  -\frac{1-R^{2}}{R^{2}}K_{0}\left(
\lambda_{3}\rho\right)  \right] \nonumber\\
&  \mp\mathbf{e}_{x}\frac{2e\mu^{\prime}}{\pi D\kappa^{2}}\int_{0}^{\infty
}d\alpha\alpha\cos\left(  \alpha x\right)  \left[  \frac{\alpha}{R^{2}}%
K_{0}\left(  \lambda_{1}\rho\right)  +\frac{\alpha}{R^{2}}K_{0}\left(
\lambda_{2}\rho\right)  -\frac{2\alpha}{R^{2}}K_{0}\left(  \lambda_{3}%
\rho\right)  \right] \nonumber\\
&  \mathbf{-e}_{\rho}\frac{2e\mu^{\prime}}{\pi D\kappa^{2}}\int_{0}^{\infty
}d\alpha\cos\left(  \alpha x\right)  \times\left[  -\frac{\left(  1+R\right)
\lambda_{1}}{2R^{2}}K_{1}\left(  \lambda_{1}\rho\right)  \right. \nonumber\\
&  \qquad\qquad\qquad\qquad\left.  -\frac{\left(  1-R\right)  \lambda_{2}%
}{2R^{2}}K_{1}\left(  \lambda_{2}\rho\right)  +\frac{\left(  1-R^{2}\right)
\lambda_{3}}{R^{2}}K_{1}\left(  \lambda_{3}\rho\right)  \right] \nonumber\\
&  \mp\mathbf{e}_{\rho}\frac{2e\mu^{\prime}}{\pi D\kappa^{2}}\int_{0}^{\infty
}d\alpha\sin\left(  \alpha x\right)  \left[  -\frac{\alpha\lambda_{1}}{R^{2}%
}K_{1}\left(  \lambda_{1}\rho\right)  -\frac{\alpha\lambda_{2}}{R^{2}}%
K_{1}\left(  \lambda_{2}\rho\right)  \right. \label{e58}\\
&  \qquad\qquad\qquad\qquad\left.  +\frac{2\alpha\lambda_{3}}{R^{2}}%
K_{1}\left(  \lambda_{3}\rho\right)  \right]  .\nonumber
\end{align}
Therefore if we set $x=0$ then
\begin{align}
\Delta\mathbf{X}\left(  0,\rho,0\right)   &  =\mathbf{e}_{x}\Delta
X_{x}(0,\rho,0)\nonumber\\
&  =\mp\mathbf{e}_{x}\frac{2e\mu^{\prime}}{\pi D\kappa^{2}}\int_{0}^{\infty
}d\alpha\frac{\alpha^{2}}{R^{2}}\left[  K_{0}\left(  \lambda_{1}\rho\right)
+K_{0}\left(  \lambda_{2}\rho\right)  -2K_{0}\left(  \lambda_{3}\rho\right)
\right]  . \label{e59}%
\end{align}
Wilson calculates the force on ion $j$ in the axial direction by further
setting $\rho=0$ in this equation, which then is given by the expression
\begin{equation}
\Delta\mathbf{X}\left(  0,0,0\right)  =\pm\mathbf{e}_{x}\frac{e\mu^{\prime}%
}{\pi D\kappa^{2}}\int_{0}^{\infty}d\alpha\frac{\alpha^{2}}{R^{2}}\ln\left(
\frac{\lambda_{1}^{2}\lambda_{2}^{2}}{\lambda_{3}^{4}}\right)  . \label{e60}%
\end{equation}
He then evaluates it by integration by parts in the form%
\begin{align}
\Delta X\left(  0,0,\theta\right)   &  =\mp\frac{e\mu^{\prime}\kappa}%
{2D}g(x),\label{e55g}\\
g(x)  &  =-\frac{1}{2x^{3}}\left[  -x\sqrt{1+x^{2}}+\tan^{-1}\left(  \frac
{x}{\sqrt{1+x^{2}}}\right)  +\sqrt{2}x-\tan^{-1}\left(  \sqrt{2}x\right)
\right]  . \label{e55w}%
\end{align}
The original formula in Wilson's thesis is different from this, but this is
the formula in the summary of his work given in the monograph\cite{harned} of
Harned and Owen. We may verify it by a different integration method as
follows. The integral in Eq. (\ref{e60}) may be written as%
\begin{align*}
I_{L}  &  \equiv\int_{0}^{\infty}d\alpha\frac{\alpha^{2}}{R^{2}}\ln\left(
\frac{\lambda_{1}^{2}\lambda_{2}^{2}}{\lambda_{3}^{4}}\right) \\
&  =\left(  \frac{\kappa}{\sqrt{2}}\right)  ^{3}\frac{1}{2\xi^{2}}\int
_{0}^{\infty}dt\frac{t^{2}}{a^{2}-t^{2}}\left[  \ln\lambda_{1}^{2}+\ln
\lambda_{2}^{2}-2\ln\lambda_{3}^{2}\right]  .
\end{align*}
Using the properties of $\lambda_{i}$ given earlier for the axial velocity
[see Eqs. (\ref{f56a})--(\ref{f57}) and Eqs. (\ref{f62})] by applying the
method of contour integrations\cite{euwien} similar to those for axial
velocity integrals, we obtain
\begin{align}
I_{L}  &  =\frac{\kappa^{3}}{4\sqrt{2}\xi^{2}}\left[  -\int_{0}^{i\sqrt
{2\left(  1+\xi^{2}\right)  }}dt\frac{t^{2}}{a^{2}-t^{2}}\left(  -\pi
i\right)  +2\int_{0}^{i}dt\frac{t^{2}}{a^{2}-t^{2}}\left(  -\pi i\right)
\right]  \quad\left(  t=\text{ complex}\right) \nonumber\\
&  =\frac{\pi\kappa^{3}}{4\sqrt{2}\xi^{2}}\left[  \int_{0}^{\sqrt{2\left(
1+\xi^{2}\right)  }}dy\frac{y^{2}}{a^{2}+y^{2}}-2\int_{0}^{1}dy\frac{y^{2}%
}{a^{2}+y^{2}}\right] \nonumber\\
&  =-\frac{\pi\kappa^{3}}{4\xi^{3}}\left[  -\xi\sqrt{\left(  1+\xi^{2}\right)
}+\sqrt{2}\xi+\frac{1}{2}\tan^{-1}\left(  2\xi\sqrt{1+\xi^{2}}\right)
-\tan^{-1}\left(  \sqrt{2}\xi\right)  \right]  . \label{55Il}%
\end{align}
Here we note that
\[
\frac{1}{2}\tan^{-1}\left(  2\xi\sqrt{1+\xi^{2}}\right)  =\tan^{-1}\frac{\xi
}{\sqrt{1+\xi^{2}}}%
\]
by the addition theorem of tangents. On inserting this integral into Eq.
(\ref{e60}), we obtain $g\left(  \xi\right)  $ as in Eq. (\ref{e55w}). The
result for $g\left(  \xi\right)  $ in Eq. (\ref{55Il}) agrees with Wilson's.

This completes the critical review of\ the calculations of the axial velocity
and ionic field and Wilson's evaluation of $f(\xi)$ and $g(\xi)$ on the basis
of Onsager's theory of conductivity. We have pointed out that taking $\rho=0$
within the integrals before they are evaluated can cause a difficulty because
of the lack of uniform convergence of some of the integrals. If $\rho\neq0$
then an alternative method of evaluating integrals must be used. An
alternative method will be discussed elsewhere\cite{euwien} that yields exact
analytical results for the integrals for all values of $\rho$. This method
recovers Wilson's results as the leading order contributions and the
correction terms are at least $O\left(  \kappa^{2}\rho^{2}\right)  $.

In any case, the axial velocity formula (\ref{e22}) given here does not appear
in the existing literature. Our analysis of the axial velocity will start from
Eq. (\ref{e22}) in the sequel\cite{euwien} to this tutorial review article.

\section{\textbf{Conductance}}

The total field strength $E$ is given by%
\begin{equation}
E=X+\Delta X_{j}(0), \label{e64}%
\end{equation}
where $\Delta X_{j}(0)$ is the field arising from the ion atmosphere
interacting with $X$. This term is called the relaxation effect, but in effect
it is a local dressed field in the modern terminology. The velocity
$\mathbf{v}_{j}$ of ion $j$ in the effective (dressed) field is then given by%
\begin{equation}
\mathbf{v}_{j}=e_{j}\omega_{j}X\left(  1+\frac{\Delta X_{j}(0)}{X}\right)
+\mathbf{v}_{x}\left(  0,0,0\right)  . \label{e65}%
\end{equation}
Here $\mathbf{v}_{x}\left(  0,0,0\right)  $ is Wilson's axial velocity. It is
useful to remark that the diffusion current (velocity) may be written relative
to $\mathbf{v}_{x}\left(  0,0,0\right)  $ as\cite{mazur}%
\begin{equation}
\mathbf{J}_{j}=\mathbf{v}_{j}-\mathbf{v}_{x}\left(  0,0,0\right)  .
\label{e66}%
\end{equation}
Then it is seen that the first term on the right of Eq. (\ref{e66}) is the
diffusion flux in the absence of a density gradient and that the so-called
electrophoresis effect is nothing but the reference velocity suitably chosen
in linear irreversible thermodynamic theory, or kinetic theory, of transport
processes. In fact, the calculation of $\mathbf{v}_{x}\left(  0,0,0\right)  $
clearly supports this interpretation. Thus inserting the results for $\Delta
X_{j}(0)$ and $\mathbf{v}_{x}\left(  0,0,0\right)  $ we obtain%
\begin{equation}
\mathbf{v}_{j}=X\left(  e_{j}\omega_{j}-\frac{e_{j}^{2}\omega_{j}\mu^{\prime
}\kappa}{DX}g(\xi)-\frac{\left\vert e_{j}\right\vert \kappa}{6\sqrt{2}\pi
\eta_{0}}f(\xi)\right)  . \label{e67}%
\end{equation}
The mobility in electrostatic units is
\begin{equation}
u_{j}=\frac{\left\vert \mathbf{v}_{j}\right\vert }{X} \label{e68}%
\end{equation}
and in practical units%
\begin{equation}
u_{j}=\frac{1}{300}\left(  \left\vert e_{j}\right\vert \omega_{j}-\frac
{e_{j}^{2}\omega_{j}\mu^{\prime}\kappa}{DX}g(\xi)-\frac{\left\vert
e_{j}\right\vert \kappa}{6\sqrt{2}\pi\eta_{0}}f(\xi)\right)  . \label{e69}%
\end{equation}
At infinite dilution $\kappa\rightarrow0$, and the limiting mobility is given
by%
\begin{equation}
u_{j}^{0}=\frac{\left\vert e_{j}\right\vert \omega_{j}}{300} \label{e70}%
\end{equation}
Since the limiting conductance is%
\begin{equation}
\Lambda_{j}^{0}=964931u_{j}^{0}, \label{e71}%
\end{equation}
we have%
\begin{equation}
\Lambda_{j}=\Lambda_{j}^{0}-\frac{e_{j}^{2}\kappa}{Dk_{B}T}\Lambda_{j}%
^{0}g(\xi)-\frac{\left\vert e_{j}\right\vert \kappa}{6\times300\sqrt{2}\pi
\eta_{0}}f(\xi). \label{e72}%
\end{equation}
With the definition of equivalent conductance%
\[
\Lambda=\Lambda_{+}+\Lambda_{-}\equiv\Lambda_{1}+\Lambda_{2},
\]
the equivalent conductance for the electrolyte is given by%
\begin{equation}
\Lambda\left(  \xi\right)  =\Lambda^{0}-\frac{z^{2}e^{2}\kappa}{Dk_{B}%
T}\Lambda^{0}g(\xi)-\frac{\left(  \left\vert e_{1}\right\vert +\left\vert
e_{2}\right\vert \right)  \kappa}{6\times300\sqrt{2}\pi\eta_{0}}f(\xi).
\label{e73}%
\end{equation}
In experiment, the relative equivalent conductance
\begin{equation}
\Delta\Lambda\left(  \xi\right)  =\Lambda\left(  \xi\right)  -\Lambda\left(
0\right)  \label{e74}%
\end{equation}
is measured and reported. We use the ratio of the relative equivalent
conductance%
\begin{equation}
\frac{\Delta\Lambda\left(  \xi\right)  }{\Lambda\left(  0\right)  }%
=\frac{\frac{z^{2}e^{2}\kappa}{Dk_{B}T}\Lambda^{0}g(0)\left[  1-\frac{g(\xi
)}{g(0)}\right]  +\frac{\left(  \left\vert e_{1}\right\vert +\left\vert
e_{2}\right\vert \right)  \kappa}{6\times300\sqrt{2}\pi\eta_{0}}f(0)\left[
1-\frac{f(\xi)}{f(0)}\right]  }{\Lambda\left(  0\right)  }. \label{e75}%
\end{equation}
This expression can be used to compute conductance and compare it with
experimental data.

\section{\textbf{Concluding Remarks}}

Charged particles in condensed phase in an external electric field are common
occurrence in physics, chemistry, and biology. It, especially, is important to
comprehend theories of transport properties of such systems and have practical
theoretical tools for them from statistical mechanics standpoints. Among the
existing theories, the theoretical development initiated a long time ago by
Onsager and his collaborators\cite{onsager,onsager1,wilson,kim,onsagerw}
provided useful theoretical models that after complex mathematical treatments
have yielded some useful well known analytical results, but his theories have
not received critical reviews by other workers that might have stimulated
studies for possible further developments in the field and other related
fields, which are many currently. His theories should be as relevant today as
they were useful when they were developed. Given the power of Onsager's
analytical theories on the subject of electrochemistry, the apparent scarcity
of critical follow-up studies of the theories\cite{chen,blum} is quite
curious. The present author sees that there seem to be considerable insights
and lessons one can draw from them only if we critically examine the theories
and benefit from them for the current topics of interest. Motivated by this
thought and their possible utilities of the theories for the currently
investigated phenomena in charged condensed matters subjected to external
electromagnetic fields, the present author has been studying his theories. In
this tutorial review article one of his early work described in Wilson's PhD
thesis has been studied critically and in detail. It is found that it has a
difficulty in one respect, but a way out of the difficulty has been found that
would improve Wilson's result for the Wien effect on equivalent conductance.
As a preparation for this line of investigation the present author has
described a tutorial review of the so-far unpublished theory up to the point
where the formal formulas for the axial velocity and ionic field are
presented. The details of derivations are given here for the formulas which
are not available in the existing scientific journal literature. The formal
formulas mentioned will be treated in full by using an alternative method of
evaluation elsewhere\cite{euwien}, which should produce exact formulas for the
electrophoretic and relaxation time effects for the Wien effect on conductance
of binary electrolytes subjected to high external electric field. The results
we obtain shows, despite the difficulty in the original evaluation of the
integrals by Wilson, his electrophoretic effect appears to be the lowest order
contribution, provided that the contributions $\left[  -eX\kappa/8\pi\eta
_{0}\left(  \kappa\rho\right)  -3eX\kappa/8\pi\eta_{0}\left(  \kappa
\rho\right)  ^{3}\right]  $ can be set equal to zero or a finite value in some
limit. What such a condition or limit would be is not clear at this point
unless the velocity expression is fully evaluated in ($x,\rho,\xi$) space.
Only then will Wilson's result would be properly understood. This aspect will
be the subject of the forthcoming work.

\bigskip

\textbf{Acknowledgement}

The present work has been supported in part by the Discovery grant from the
Natural Sciences and Engineering Council of Canada.

\end{document}